\documentclass[letterpaper, 10 pt, conference]{ieeeconf}
\IEEEoverridecommandlockouts 
\usepackage{graphicx} 
\usepackage{cite}
\usepackage{amsmath}
\usepackage{amsfonts}
\usepackage{dsfont}
\let\labelindent\relax
\usepackage{enumitem,kantlipsum}
\usepackage{subcaption}
\usepackage{bm}
\usepackage{multirow}
\usepackage{color}

\usepackage{booktabs}  
\usepackage{threeparttable}
\usepackage[english]{babel}

\usepackage[linesnumbered,ruled,vlined]{algorithm2e}

\usepackage{algcompatible}
\SetKwComment{Comment}{/* }{ */}

\newtheorem{rem}{Remark}

\usepackage{amssymb}
\usepackage{xcolor}
\date{} 

\title{\LARGE \bf Optimal Safe Sequencing and Motion Control \\ for Mixed Traffic Roundabouts}

\author{Yingqing Chen and Christos G. Cassandras 
}

\begin{document}
\maketitle
\thispagestyle{empty}
\pagestyle{empty}
\SetKwInput{KwInput}{Input}                
\SetKwInput{KwOutput}{Output} 

\begin{abstract} 

This paper develops an Optimal Safe Sequencing (OSS) control framework for Connected and Automated Vehicles (CAVs) navigating a single-lane roundabout in mixed traffic, where both CAVs and Human-Driven Vehicles (HDVs) coexist. The framework jointly optimizes vehicle sequencing and motion control to minimize travel time, energy consumption, and discomfort while ensuring speed-dependent safety guarantees and adhering to velocity and acceleration constraints. This is achieved by integrating (a) a Safe Sequencing (SS) policy that ensures merging safety without requiring any knowledge of HDV behavior, and (b) a Model Predictive Control with Control Lyapunov Barrier Functions (MPC-CLBF) framework, which optimizes CAV motion control while mitigating infeasibility and myopic control issues common in the use of Control Barrier Functions (CBFs) to provide safety guarantees. Simulation results across various traffic demands, CAV penetration rates, and control parameters demonstrate the framework's effectiveness and stability.

\end{abstract}




\section{Introduction}
Urban congestion has reached critical levels in recent decades, with significant impacts on travel time, pollution, and fuel consumption \cite{chang2017there}. 
This congestion primarily stems from conflict areas such as intersections, roundabouts, merging roadways, and speed reduction zones \cite{rios2016automated}. The emergence of Connected and Automated Vehicles (CAVs) along with real-time communication infrastructure \cite{liu2023} offers promising solutions for achieving smoother traffic flow and lower fuel consumption \cite{campi2023roundabouts} through enhanced information utilization and precise trajectory design.


While most research has focused on controlling CAVs within a single \emph{Control Zone} (CZ) that encompasses a specific conflict area such as merging roadways \cite{milanes2010automated,xiao2021decentralized},  lane changing \cite{armijos2024cooperative,li2023cooperative}, and unsignalized intersections \cite{bichiou2018developing,zhang2019decentralized}, the transition to multiple interconnected CZs presents unique challenges \cite{xu_scaling_2023}.
While an isolated CZ assumes that incoming vehicles satisfy all constraints, interconnected CZs often lead to constraint violations as vehicles transition between zones. Moreover, optimizing control for a single CZ without considering subsequent zones may result in excessive control effort, increased congestion, or even blocked traffic.

A typical case in point of interconnected CZs arises in a roundabout, a configuration which is attractive to traffic control because its geometry enhances safety by promoting lower speeds, better visibility, increased reaction time for drivers, and resulting in less severe crashes when they do occur \cite{rodegerdts2010roundabouts}. Though roundabouts offer many such benefits, their effective management requires more careful planning, since the compact geometry with multiple closely-spaced \emph{Merging Points} (MPs) requires tighter control and more stringent safety constraint satisfaction.
Various approaches, both model-based \cite{bakibillah_bi-level_2021,bichiou_developing_2019,zhao_optimal_2018,xu2021decentralized} and learning-based \cite{garcia_cuenca_autonomous_2019,capasso_intelligent_2020}, have been explored for roundabout control under full CAV scenarios.

However, the assumption of 100\% CAV penetration remains impractical. Mixed traffic scenarios, where CAVs and Human Driven Vehicles (HDVs) coexist, present a more realistic near-term solution. Even with a small proportion of CAVs, these scenarios show significant improvements over fully HDV-based systems \cite{mohebifard2022trajectory,ferrarotti2024autonomous}. Yet, managing mixed traffic introduces additional challenges beyond those encountered in fully autonomous systems. Human driving behavior is inherently unpredictable, complicating efforts to maintain both safety and efficiency. 

Current approaches often employ game theory to model HDV-CAV interactions, such as entropic risk minimization \cite{wang2020game}, bi-level Stackelberg optimization \cite{burger2022interaction}, and receding horizon game-theoretic planning \cite{wang2021game}. Other studies model human behavior using optimization frameworks \cite{sadigh2018planning,burger2022interaction} or predictive models \cite{mohebifard2022trajectory,zhang2023social}. While these approaches offer valuable insights, they rely on strong human behavioral assumptions that may not generalize across different drivers, situations, or traffic configurations, particularly in roundabouts where multiple MPs create complex decision-making environments. In addition, coordinating different kinds of vehicles across interconnected CZs increases computational complexity, potentially compromising real-time performance in practical applications.

To address these challenges, motivated by findings that human drivers following CAVs exhibit more stable behavior \cite{mahdinia2021integration} and that CAV platoons before entering a MP can improve throughput \cite{zhuo2023evaluation}, we propose an Optimal Safe Sequencing (OSS) framework that ensures safety and efficiency in mixed traffic roundabouts without relying on strong assumptions about HDV behavior. 
When potential HDV-CAV interactions arise, the framework ensures \emph{safety} by enforcing a Safe Sequence (SS) policy which induces desirable HDV behavior through its preceding CAV (if present), thereby preventing HDV-CAV conflicts at MPs. Specifically, the policy prohibits CAVs from merging ahead of HDVs when the distance between them is small.
Such a policy mitigates the risk of safety violations caused by the unpredictable merging behavior of HDVs when encountering conflicting CAVs.
On the other hand, \emph{optimality} is achieved by integrating the SS policy with a Model Predictive Control - Control Lyapunov Barrier Function (MPC-CLBF) framework \cite{chen2024optimal}.
This framework combines MPC with constraints formulated as Control Barrier Functions (CBFs) and Control Lyapunov Barrier Functions (CLBFs), leveraging their forward invariance, finite convergence properties, and computational efficiency. 
The combined framework ensures safety while jointly optimizing vehicle sequencing and motion control for improved traffic flow, energy efficiency, and driver comfort.
 
The main contributions of this paper are as follows:
\begin{itemize}
    \item [1.] We propose a SS policy that ensures safety in mixed traffic roundabouts without requiring strong 
    assumptions about HDV behavior.
    \item [2.] We integrate the SS policy with an MPC-CLBF to form the OSS framework, which jointly optimizes sequencing and motion control to balance speed, energy efficiency, and comfort while ensuring safety.
    \item [3.]We conduct extensive simulations under varying traffic demand structures, CAV penetration rates, sequencing policies, and receding horizon parameters. Our results demonstrate that the SS policy combined with MPC-CLBF outperforms naive yielding policies and remains robust across different traffic conditions. Additionally, we show that higher CAV penetration rates further enhance driving efficiency under our proposed policy.
\end{itemize}

The paper is organized as follows. In Section \ref{Sec:Problem Formulation}, the mixed traffic roundabout problem is formulated as an optimal control problem with safety constraints. In Section \ref{Sec:Optimal Resequencing}, a joint optimal safe sequencing and motion control approach is developed. Simulation results are presented in Section \ref{Sec: Simulation} showing effectiveness of our controller under different traffic demand structure, CAV penetration rate and parameter settings. Finally, in Section \ref{sec: conclusion} we provide conclusions and future research.

\section{Problem Formulation}\label{Sec:Problem Formulation}

\begin{figure}
    \centering
    \includegraphics[width=0.99\columnwidth]{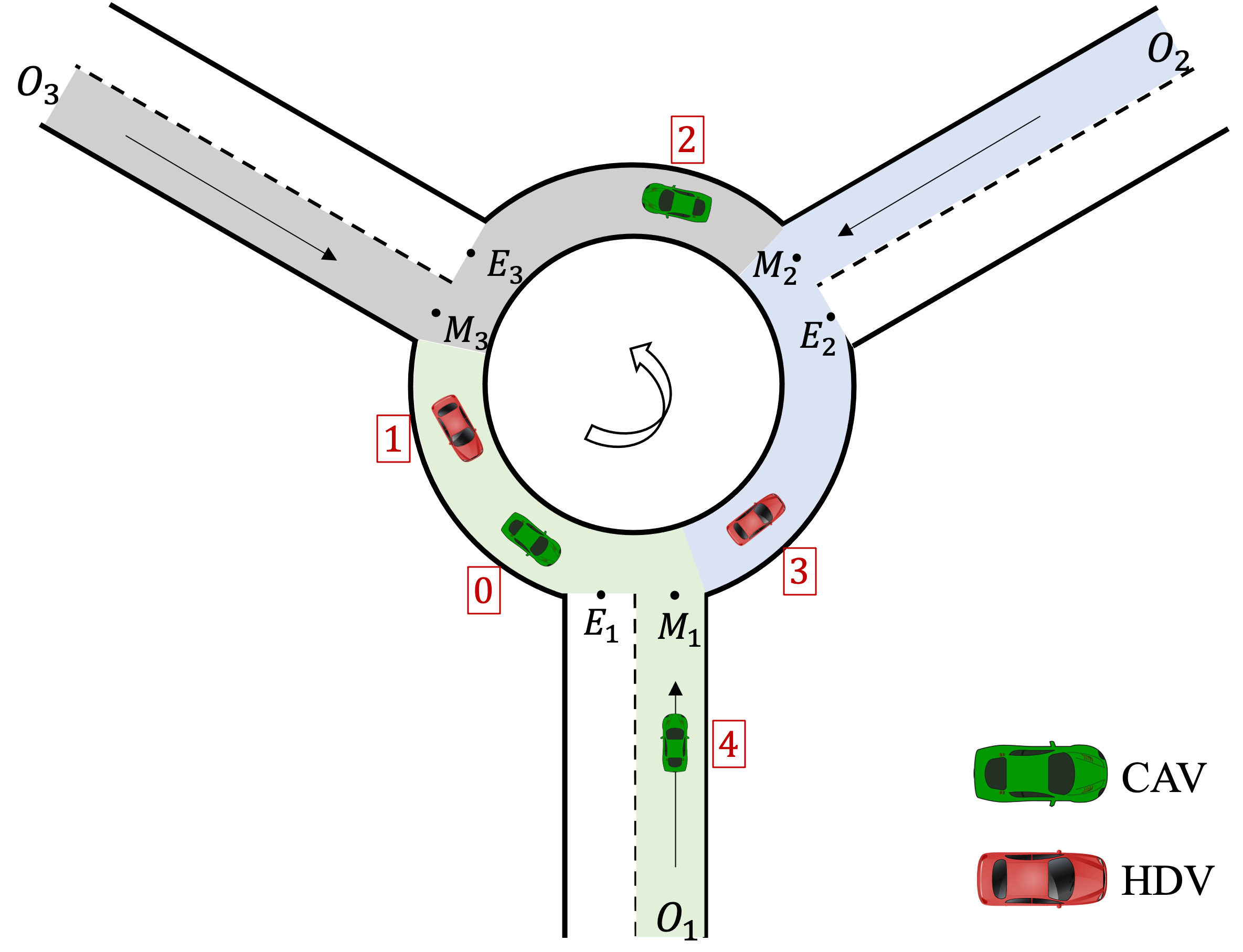}
    \caption{A roundabout with 3 entries }
    \label{fig:roundabout}
\end{figure}

We consider a single-lane roundabout with $N$ entry and $N$ exit points where $N\geq 2$. For simplicity we limit our analysis here to $N=3$, as depicted in Fig. \ref{fig:roundabout}, with straightforward extensions to $N>3$. We assume that each road segment has a single lane (extensions to multiple lanes are possible following natural extensions that have been already applied in merging\cite{xiao2020decentralized}  and in signal-free intersections \cite{xu2022general}).

In this paper, we consider traffic to consist of both HDVs and CAVs (red and green vehicles, respectively, in Fig. \ref{fig:roundabout}) which randomly enter the roundabout from three different origins $O_1$, $O_2$ and $O_3$ and have randomly assigned exit points $E_1$, $E_2$ and $E_3$.
We assume all vehicles move in a counterclockwise way in the roundabout. 
The entry road segments are connected with the circle at the three Merging Points (MPs) labeled $M_1$, $M_2$ and $M_3$, where vehicles may potentially laterally collide with each other. Associated with each MP, we partition the roundabout into three Control Zones (CZs), labeled $CZ_1$, $CZ_2$ and $CZ_3$; these are shaded in green, blue, grey respectively in Fig.\ref{fig:roundabout}. Each CZ includes two road segments with traffic headed towards the corresponding MP: one is the straight entry road indexed by $1$, and the other is the curved road segment within the roundabout, indexed by $0$. 


For $CZ_k$, we set the length of the straight entry road and inside curve road as $L_k^1$ and $L_k^0$ respectively, $k\in \{1,2,3\}$. 
Also, we define $c_i \in \{0,1\}$ as the road segment where vehicle $i$ is currently located, where $c_i=1$ indicates vehicle $i$ is at an entry road, otherwise $c_i=0$.
The full sequence of MPs a vehicle must go through can be determined by its entry and exit points.

Let $S(t)$ be the set of indices of vehicles present in the roundabout at time $t$. 
This set comprises indices corresponding to CAVs and HDVs, denoted by $S^c(t)$ and $S^h(t)$, respectively, such that $S^c(t)\cup S^h(t) = S(t)$ and $S^c(t) \cap S^c(t) = \emptyset$.
We also partition $S(t)$ into subsets based on their current CZ and define a \emph{merging group} as a set of vehicles present at the same CZ at time $t$. Let $S_k(t)$, $k \in \{1,2,3\}$, be the set of vehicle indices (from $S(t)$) within the merging group of $CZ_k$ at time $t$, where $S_1(t)\cap S_2(t) \cap S_3(t) = \emptyset$, $S(t)=S_1(t)\cup S_2(t) \cup S_3(t)$. Each subset $S_k(t)$ can be further partitioned into $S_k^C(t)$ and $S_k^H(t)$ respectively, representing the indices of CAVs and HDVs inside $CZ_k$ respectively. 

The dynamics and control policy of the HDV are unknown in this case. The vehicle dynamics for each CAV $i\in S^c(t)$ along the road segment to which it belongs take the form
\begin{equation}
\left[
\begin{array}
[c]{c}%
\dot{x}_{i}(t)\\
\dot{v}_{i}(t)
\end{array}
\right]  =\left[
\begin{array}
[c]{c}%
v_{i}(t)\\
u_{i}(t)
\end{array}
\right]  \label{VehicleDynamics}%
\end{equation}
where $x_{i}(t)$ denotes the distance from the origin of the road segment where vehicle $i$ currently locates, $v_{i}(t)$ denotes the velocity, and $u_{i}(t)$ denotes the control input (acceleration). We also define $d_i(t)$ as the total distance vehicle $i$ has traveled from the origin of its route; this can be derived directly by adding to $x_i(t)$ the lengths of all roads passed. 

In this paper, we focus on optimal sequencing and longitudinal motion control that guarantee safety under mixed traffic, while disregarding lateral vehicle motion, which would not affect the safety and efficiency metric in our single lane scenario. We assume the presence of a steering controller that is applied separately for each vehicle to maintain their position within the lane boundaries, therefore we solely address control for longitudinal displacement along the road. However, the lateral vehicle offset control can be addressed by employing more detailed models, such as the bicycle model in \cite{polack2017kinematic} or the 7-state model in \cite{rucco2015efficient}, within the same framework presented in this paper. In fact, our previous work\cite{xu2022decentralized} has already demonstrated the effectiveness of handling the lateral offset using dynamics with respect to a reference trajectory, which measures the along-trajectory distance and the lateral distance of the vehicle's Center of Gravity (CoG) from the closest point on the reference trajectory. The same structure can be applied to our current framework as well.


In this paper, we assume that each vehicle type  (HDVs or CAVs) and its state can be estimated using sensing equipment in the road network available to a roadside (RSU) coordinator unit. We further assume that only CAVs are equipped with connectivity, enabling them to communicate their states, control inputs, and roundabout exit intentions to the coordinator; HDVs have no connectivity capability.
Based on the information available for CAVs and HDVs, the motion control of each CAV can be determined by solving an optimal control problem as presented next.

\subsection{Optimal Control Problem}\label{subsec:optimal control problem}
We consider three objectives for each CAV subject to four constraints, as detailed next.

$\emph{Objective 1:}$ Minimize the travel time $J_{i,1} = t^f_i - t_i^0$ where $t_i^0$ and $t^f_i$ are the times CAV $i$ enters and exits the roundabout.

$\emph{Objective 2:}$ Minimizing energy consumption:
\begin{equation}
\label{energycost}
\setlength{\abovedisplayskip}{2pt}\setlength{\belowdisplayskip}{2pt}
J_{i,2}=\int_{t_{i}^{0}}^{t_{i}^{f}}C(u_{i}(t))dt,
\end{equation}
where $C(\cdot)$ is a strictly increasing function of its argument. Since the energy consumption rate is a monotonic function of the acceleration, we adopt this general form to achieve energy efficiency.

$\emph{Objective 3:}$
Minimizing centrifugal discomfort:
\begin{equation}
\label{centrifugalcost}
\setlength{\abovedisplayskip}{2pt}\setlength{\belowdisplayskip}{2pt}
J_{i,3}=\int_{t_{i}^{0}}^{t_{i}^{f}}\kappa_i(d_i(t))v_i^2(t)dt,
\end{equation}
where $\kappa_i(d_i(t))$ is the curvature of the road within route of CAV $i$ at position $d_i(t)$. As the aim is to minimize the centrifugal force of the vehicle, the curvature $\kappa_i(d_i(t))$ has the form of $\frac{1}{r_i(d_i(t))}$, where $r_i(d_i(t))$ is the radius of the road at position $d_i(t)$ .

\begin{figure}
    \centering
    \includegraphics[width=0.6\columnwidth]{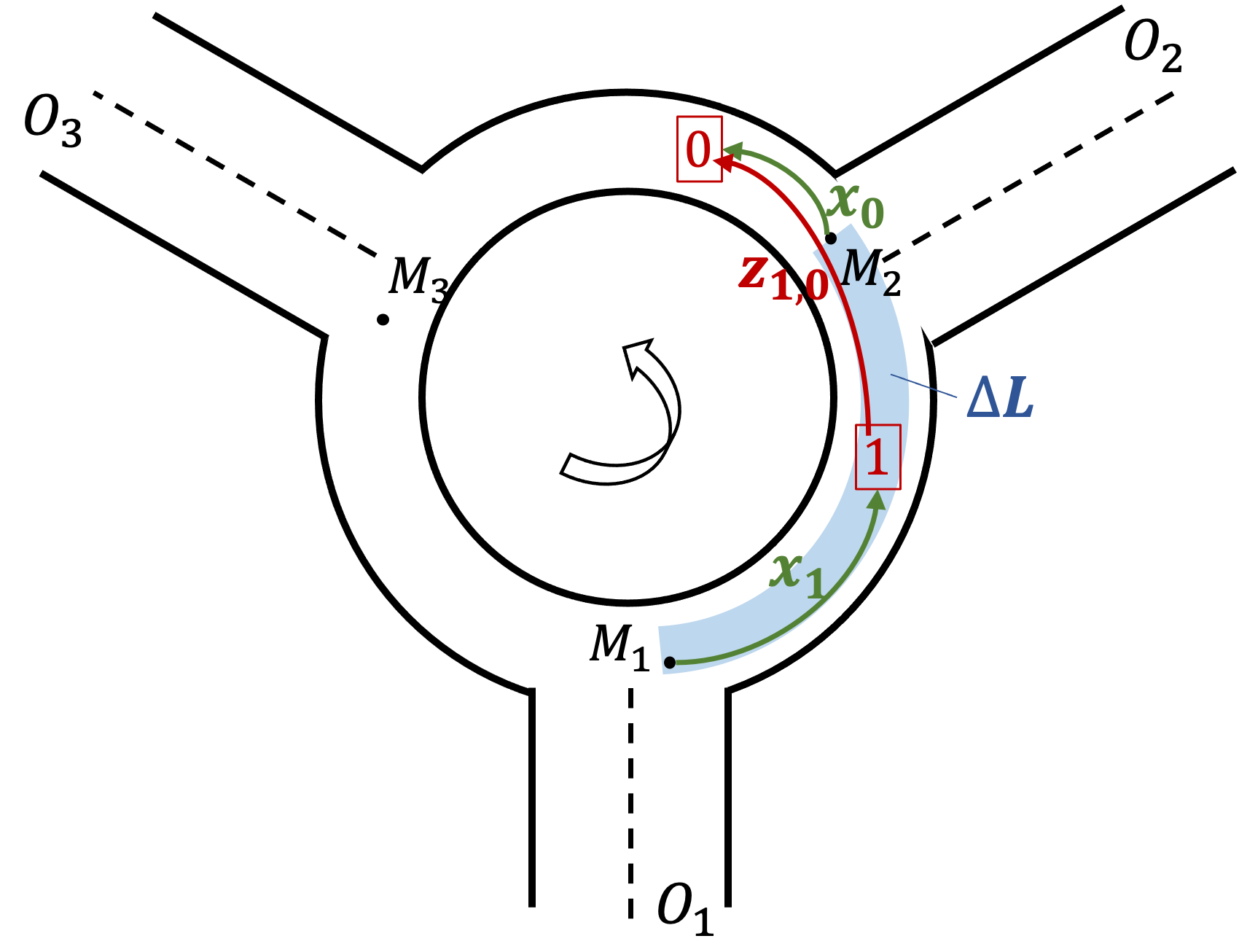}
    \caption{Illustration of $z_{i,i_{p}}$ definition for $i=1$, $i_p=0$}
    \label{fig:distance_gap}
\end{figure}

\emph{Constraint 1 (Rear end safety constraints)}: 
We define $i_{p}$ as the index of the vehicle which physically immediately precedes vehicle $i$ in the roundabout (if one exists), regardless of whether they are in the same CZ.  We define the distance between $i$ and $i_p$ as $z_{i,i_{p}}(t):=\bar x_{i_{p}}(t)-x_{i}(t)$, where $\bar{x}_{i_p}(t)$ is the position of vehicle $i_p$ relative to the starting point of the (possibly different) road segment containing vehicle $i$. This position is given by 
$\bar{x}_{i_p}(t) = x_{i_p}(t) + \Delta L$, where $\Delta L$ denotes the distance between the starting points of the two road segments. As illustrated in Fig. \ref{fig:distance_gap} where $i = 1$, $i_p=0$ (indicating that vehicle 0 immediately precedes vehicle 1), the distance gap is $z_{1, 0} = x_0 + \Delta L - x_1$ where $\Delta L$ corresponds to the the distance between $M_1$ and $M_2$. We require that
\begin{equation}
\setlength{\abovedisplayskip}{2pt}\setlength{\belowdisplayskip}{2pt}
z_{i,i_{p}}(t)\geq\varphi v_{i}(t)+\delta,\text{ \ }\forall t\in\lbrack
t_{i}^{0},t_{i}^{f}], \label{equ: rear-end safety const}
\end{equation}
where 
$v_{i}(t)$ is the speed of CAV $i\in S^c(t)$ and $\varphi$ denotes the reaction time (as a rule, $\varphi=1.8$s is used, e.g., \cite{Vogel2003}). If we define $z_{i,i_{p}}$ to be the distance from
the center of CAV $i$ to the center of vehicle $i_{p}$, then $\delta$ is a constant determined by the length of these two vehicles (generally dependent on $i$ and $i_{p}$ but taken to be a constant for simplicity).

\emph{Constraint 2 (Safe merging constraint)}:

We define $i_m$ as the index of the last vehicle that precedes vehicle $i$ (if one exists) at the next MP, $M_k$, within a given sequence. Vehicle $i_m$ must be in the same CZ but a different road segment than that of vehicle $i$, i.e., $c_i \ne c_{i_m}$. 
Let $t_{i_m}^k, k \in\{1, 2, 3\}$ be the arrival time of vehicle $i_m$ at $M_k$. Denote the lengths of the roads where CAV $i$ and vehicle $i_m$ are currently located as $L_i$ and $L_{i_m}$ respectively.
The distance between $i_m$ and $i$, given by $z_{i,i_m} (t)\equiv (L_i - x_i(t)) - (L_{i_m}-x_{i_m} (t))$, is constrained by
\begin{equation}
z_{i,i_m}(t_{i_m}^{k})\geq\varphi v_{i}(t_{i_m}^{k
})+\delta,  ~~\forall i\in S(t), ~k\in \{1,2,3\} \label{equ: Merging safe}%
\end{equation}

\emph{Constraint 3 (Lateral Safety Constraint)}: The moment generated by the centrifugal force needs to be smaller than the one generated by gravity in order to avoid rollover:
\begin{equation}
\kappa_i(d_i(t))\cdot v_i^2(t)\cdot h \leq w_h\cdot g \label{equ: lateral safe}
\end{equation}
where $h$ is the height of the CAV, $w_h$ is its half width (for simplicity, both assumed the same for all CAVs) and $g$ is the gravity constant.

\emph{Constraint 4 (Vehicle limitations)}: There are constraints on the speed and acceleration for each $i\in S^c(t)$, i.e.,
\begin{align}
v_{i,\min} \leq v_i(t)\leq v_{i,\max}, ~~\forall t\in[t_i^0,t_i^f],\\
u_{i,\min}\leq u_i(t)\leq u_{i,\max}, ~~\forall t\in[t_i^0,t_i^f],  \label{equ:VehicleConstraints}%
\end{align}
where $v_{i,\max}>0$ and $v_{i,\min}\geq0$ denote the maximum and minimum speed
allowed in the roundabout, while $u_{i,\min}<0$ and $u_{i,\max}>0$ denote the minimum
and maximum control input for CAV $i$, respectively. We further assume common speed limits dictated by the traffic rules at the roundabout, i.e. $v_{i,\min} = {v_{\min}}, v_{i,\max} = v_{\max}$.

\textbf{Problem 1: } Our goal is to determine a control law to achieve objectives 1-3 subject to constraints 1-4 for each CAV $i\in S^c(t)$ governed by the dynamics (\ref{VehicleDynamics}). 
Thus, we combine Objectives 1-3 by constructing a convex combination
as follows: {
\begin{equation}
\setlength{\abovedisplayskip}{2pt}\setlength{\belowdisplayskip}{2pt}
\begin{aligned}J_i=\left(\alpha_1 J_{i,1} + \alpha_2 \frac{J_{i,2}}{\frac{1}{2}\max \{u_{i,\max}^2, u_{i,\min}^2\}} + \alpha_3 \frac{J_{i,3}}{\kappa_{\max}v_{\max}^2}\right) \end{aligned}\label{eqn:energyobja}%
\end{equation}
}where $\alpha_1, \alpha_2, \alpha_3 \in [0,1], \alpha_1 + \alpha_2 + \alpha_3 = 1$, $J_{i,2}$ and $J_{i,3}$ are properly normalized. 
Note that these weights are determined by each CAV according to its owner’s relative preferences among time, energy, and comfort; they are set upon a CAV’s arrival and remain unchanged during the CAV’s trip through the roundabout. 

In what follows we simply choose in $J_{i,2}$ the quadratic function $C_i(u_i(t)) =\frac{1}{2} u_i^2(t)$.
Then, by defining $\beta_1:=\frac{\alpha_1\max\{u_{i,\max}%
^{2},u_{i,\min}^{2}\}}{2 \alpha_2}$, $\beta_2 := \frac{\alpha_3 \max\{u_{i,\max}%
^{2},u_{i,\min}^{2}\}}{2\alpha_2 \kappa_{\max}v_{\max}^2}$, we can rewrite this minimization problem as: 
{\small
\begin{equation}
\setlength{\abovedisplayskip}{1pt}\setlength{\belowdisplayskip}{1pt}J_{i}%
(u_{i}(t)):=\int_{t_{i}^{0}}^{t_{i}^{f}%
}(\beta_1 + \frac{1}{2}u_{i}^{2}(t) + \beta_2\kappa_i(d_i(t))v_i^2(t))dt,\label{eqn:obj1}%
\end{equation}
}where $\beta_1\geq0$ and $\beta_2\geq0$ are weights factors, {subject to
(\ref{VehicleDynamics}), (\ref{equ: rear-end safety const})-(\ref{equ:VehicleConstraints}) given $t_{i}^{0},v_{i}(t_{i}^{0})$.}

There are three main challenges in this problem: 

(1) Solving this problem for any CAV $i$ requires knowledge of $i_p$ and $i_m$ in (\ref{equ: rear-end safety const})-(\ref{equ: Merging safe}) whose assignment is dynamically changing, especially when a vehicle changes CZs. 

(2) Even if $i_p$ and $i_m$ are determined for each CAV, such information remains unknown to HDVs. 
We assume that HDVs can maintain rear-end safety and do not overtake their preceding vehicles, thereby preserving the fixed $i_p$ assignments. However, HDVs can disrupt the merging sequence defined by $i_m$ assignments of CAVs.
To capture this disruption, we define $i_m^-$ as the index of the first vehicle that follows vehicle $i$ (if one exists) at the next MP within a given sequence, such that $c_i \neq c_{i_m^-}$. In other words, the relationship between the vehicle pair ($i$, $i_m$) mirrors that of $(i_m^-, i)$. 
To ensure safe merging at any MP, CAV $i$ must maintain sufficient distances to both $i_m$ and $i_m^-$ (i.e., adequate distance gaps $z_{i,i_m}(t_{i_m}^k)$ and $z_{i_m^-,i}(t_{i}^k)$)), where CAV $i$ knows the vehicle types of both $i_m^-$ and $i_m$.
For the vehicle pair ($i$, $i_m$), the safety is guaranteed by the safe merging constraint \eqref{equ: Merging safe} of $i$ even when $i_m$ is an HDV. 
For the vehicle pair $(i_m^-, i)$, if $i_m^-$ is a CAV, the merging safety is naturally ensured by the safe merging constraint \eqref{equ: Merging safe} applied to CAV $i_m^-$. However, if $i_m^-$ is an HDV, and CAV $i$ plans to merge ahead of this HDV, an adequate distance gap $z_{i_m^-,i}(t_{i}^k)$ cannot be ensured due to the unpredictable behavior of HDV $i_m^-$. For example, HDV $i_m^-$ might unexpectedly accelerate and merge ahead of CAV $i$, disrupting the vehicle sequence($i_m^-,i,i_m$) and compromising overall merging safety. 

Such sequence randomness and safety violations cannot be resolved by simply adding a constraint to $z_{i_m^-,i}(t_{i}^k)$, as the dynamics and control policies of HDVs are unknown. 
This indicates that the vehicle sequence ($i_m^-,i,i_m$) is fundamentally unsafe when $i_m^-$ is an HDV.
This necessitates the development of alternative safety strategies to address conflicts from unpredictable HDV behavior, accounting for uncertainty while providing robust safety guarantees despite limited information about human driver responses during merges.

(3)	Solving \textbf{Problem 1} is computationally prohibitive for real-time applications.

To address these challenges, our approach is to determine an \emph{optimal safe sequence} for each merging group in a CZ under mixed traffic conditions while simultaneously deriving the corresponding optimal control $u_i(t)$, $i\in S^c(t)$ efficiently under such safe sequencing. 
The control $u_i(t)$ is designed to ensure safety during interactions with other vehicles, while the safe sequence provides an additional layer of protection to mitigate potential conflicts arising from the unpredictable behavior of HDVs.


\section{Optimal Safe Sequencing and Motion Control}\label{Sec:Optimal Resequencing}
Our objective is to assign each CAV (a) an appropriate safe order within its merging group with respect to the next MP that remains robust to HDV behavior, and (b) associated motion control so as to minimize the total traveling time, energy and centrifugal discomfort as in \eqref{eqn:obj1} while satisfying the safety constraints (\ref{equ: rear-end safety const})-(\ref{equ: lateral safe}).

To solve this problem, we use a joint process of optimal safe sequencing and motion control that comprises three steps: 
(a) identify all feasible safe sequences that remain robust against the random behavior of HDVs and can be followed by all CAVs in the merging group when crossing the MP,
(b) evaluate each candidate feasible safe sequence using the MPC-CLBF framework, and 
(c) based on the first two steps, determine the optimal safe sequence and the associated control for each CAV. 
These three steps are detailed next.

\subsection{Find Feasible Safe Sequence Set}
CAVs not located at their final CZ can conflict with CAVs from other inbound road segments heading to the shared next MP. This creates flexibility in their relative merging order, which we define as different \emph{sequences}.
A sequence is a prioritized list of indices for all vehicles currently within the same CZ, ordered by projected CZ exit times (either progressing to the next CZ or leaving the roundabout). 
Given the single lane assumption (which excludes overtaking), any \emph{feasible} sequence must maintain relative on-road precedence.
In addition to feasibility, any selected sequence must also be \emph{safe}.

\textbf{Safe Sequence.}
The concept of Safe Sequence (SS) was first introduced in \cite{sabouni_merging_2023} to address merging challenges under mixed traffic. 
Similarly in our case, to exclude the effect of HDV behavior from sequence determination at the MPs, we ensure that \emph{no CAV merges ahead of an HDV}. For example, in Fig.~\ref{fig:roundabout}, the feasible sequences for $CZ_1$ includes $[4, 0, 1]$, $[0, 4, 1]$ and $[0, 1, 4]$. Here, CAV 4 can merge ahead of CAV 0 or HDV 1, or merge behind all vehicles. By enforcing the rule that \emph{no CAV merges ahead of an HDV}, we eliminate sequence $[0, 4, 1]$, leaving two valid options: (a) CAV 0 decelerates and makes room for CAV 4 to merge ($[4, 0, 1]$), or (b) CAV 4 decelerates and merges after HDV 1 ($[0, 1, 4]$).
This strategy promotes cooperation between CAVs and prevents situations where an HDV $i_m^-$ accelerates to merges ahead of a CAV $i$. As a result, it mitigates the risk of safety violations due to the unpredictable behavior of HDVs and enhances the robustness of the merging sequence.



However, this approach can be overly conservative in certain cases. If HDV $i_m^-$ is sufficiently far from the MP relative to CAV $i$, requiring the CAV to yield imposes unnecessary restrictions. To prevent this, we introduce a distance threshold $g_j$ that permits a CAV to merge ahead of an HDV $j$ if their relative distance to the next MP exceeds this threshold, ensuring that the HDV’s behavior does not influence the CAV’s trajectory. Specifically, 
HDV $j$ does not interact with or influence the behavior of CAV $i$ if the following condition is satisfied:
\begin{equation}\label{eqn: HDV merging distance constraint}
    z_{j,i}(t) - \varphi (v_j(t) - v_i(t)\cdot \frac{x_i(t)}{L_i(t)}) \geq g_j
\end{equation}
where $L_i(t)$ denotes the length of the road segment where CAV $i$ is currently located. If this constraint is violated, merging ahead of HDV $j$ is not allowed, and CAV $i$ must either yield to $j$ or merge ahead of another CAV to enforce a safe sequence.
Notably, the left-hand side of \eqref{eqn: HDV merging distance constraint} decreases when CAV $i$ is farther from the next MP, as indicated by a smaller $x_i(t)$. This makes it more likely for the constraint to be violated, thereby promoting yielding to HDV $j$. The reasoning is that when the merging point is farther away, both CAV $i$ and HDV $j$ have a longer time before reaching it, increasing the uncertainty in HDV $j$’s behavior. This added uncertainty raises the safety risk if merging ahead of HDV $j$. Additionally, the extended time horizon allows CAV $i$ more flexibility to adjust its trajectory smoothly, reducing the likelihood of abrupt control changes.

\begin{rem}
Intuitively, the distance threshold $g_j$ should account for the driving behavior of HDV $j$. An aggressive HDV requires a larger safety gap, while a conservative HDV allows a CAV to merge with a smaller gap without compromising safety. Research has shown that drivers can distinguish between aggressive and conservative behaviors based on past trajectories \cite{chandra2022game}. Thus, it is reasonable to infer an HDV's aggressiveness from historical driving patterns or recent trajectories observed by the coordinator within the CZ. 
When such behavioral data is available, the distance threshold can be calibrated to respond appropriately to different HDV types, optimizing space allocation for safe sequencing. To formalize this approach, we define an aggressiveness parameter $a_j \in [-1,1]$ for each HDV $j \in S^h(t)$, which is derived from past state and control trajectories, where higher values indicate greater aggressiveness. 
The distance threshold is then formulated as an increasing function of $a_j$, i.e., $g_j = g(a_j)$,  ensuring that more aggressive HDVs require a larger safety gap. We propose a cubic function:
\begin{equation} \label{eqn: HDV safety threshold}
    g(a_j) = \delta_j + \eta \cdot a_j^3
\end{equation}
where $\delta_j$ represents the baseline threshold for an average-behavior HDV $j$ ($a_j=0$), and $\eta$ adjusts sensitivity to aggressiveness.
The cubic form ensures that extreme behaviors (highly aggressive or conservative) receive extra caution, while responses to near-average behavior remain stable.
$\hfill\blacksquare$
\end{rem}

 \textbf{Feasible Safe Sequences}
Using the concepts above, we generate feasible safe sequences through the following steps: 
\begin{itemize}
    \item Generate feasible sequences: List all possible sequences for a merging group while preserving on-road precedence.
    \item Check safety constraint: For each sequence, verify whether any CAV $i$ is placed ahead of an HDV $j$ from a different road segment ($c_i \neq c_j$) without satisfying the relative distance constraint \eqref{eqn: HDV merging distance constraint}.
    \item Filter unsafe sequences: Remove any sequence that violates the above constraint, resulting in a set of \emph{feasible safe sequences}.
\end{itemize}
 We denote the set of all feasible safe sequences for $CZ_k$ at time $t$ as $F_k(t)$. 
For the same example of $CZ_1$ in Fig.~\ref{fig:roundabout},
if $z_{1, 4}(t)-\varphi (v_1(t) - v_4(t)\cdot \frac{x_4(t)}{L_4(t)}) < g_4$, 
the feasible safe sequences for $S_1(t)$ are $F_1(t) = \{[0,1,4], [0,4,1],[4,0,1]\}$. Otherwise, $F_1(t) = \{[0,1,4], [4,0,1]\}$.

Note that our SS strategy ensures safety even with \emph{unknown dynamics} for the HDVs. When adopting a specific feasible safe sequence, although we cannot directly control the behavior of HDVs, we indirectly influence their actions through the behavior of CAVs. In the previous example, when choosing sequence $[4,0,1]$ , CAV 0 decelerates to create space for CAV 4 to merge ahead, and compel HDV 1 to decelerate at the same time, thus excluding HDV 1 from potential conflict at the next MP. This approach ensures that, despite HDVs being part of the merging process, their unpredictable behavior does not affect control decisions, thereby maintaining safety and sequence integrity.

\textbf{Assign $i_p, i_m$.} 
Each sequence determines the $i_p$ and $i_m$ assignments for each vehicle $i$.
For each feasible safe sequence $\bm f\in F_k(t)$,
we first partition it into two subsequences, $\bm f_0$ and $\bm f_1$, based on the $c_i$ value of each vehicle, while preserving the original order: $i \in \bm f$ is added to $\bm f_0$ if $c_i=0$, otherwise it is added to $\bm f_1$.
Then, $i_m$ is assigned to the closest vehicle index that precedes $i$ in $\bm f$ and belongs to the subsequence which  $i$ does not belong to (i.e., $c_i \neq c_{i_m}$). For example, for $i=4$, under sequence $\bm f=[0,1,4]$, $\bm f_0=[0,1]$, $\bm f_1=[4]$ and $4_m=1$.

Regarding $i_p$, there are two cases. First, if $i$ does not rank first in its subsequence, its $i_p$ is the index immediately prior to $i$ in this subsequence (e.g., for $\bm f=[0,1, 4]$, $1_p=0$). If, on the other hand, it ranks first, then there are two cases to consider:
(a) If $i$ is located at its final CZ (indicating that it is leaving the roundabout), then $i_p$ does not exist (e.g., $0_p$ does not exist when CAV 0 leaves at $E_1$
).  
(b) Otherwise, we need to determine $i_p$ by finding a vehicle located in the next CZ which immediately precedes vehicle $i$. If the current CZ is $k$, the next one is denoted by $k^+ \equiv k+1 \pmod{3}$. If there exists some vehicle $j \in S_{k^+}(t)$ such that $c_j=0$ and $j$ does not exist in the $i_p$ column of $S_{k^+}(t)$ (which indicates vehicle $j$ is the last one on its road segment), 
we set $i_p = j$ (e.g., $4_p=3$ in Fig. \ref{fig:roundabout}). If not, we move to the next CZ and check CAV $i$ following the same logic until we return to the initial CZ, indicating that no $i_p$ exists. 

\subsection{Evaluate a Feasible Safe Sequence}
After determining the feasible safe sequence set $F_k(t)$ for each merging group in $CZ_k, k\in\{1,2,3\}$, we need to find the optimal sequence. To do so we need to evaluate each sequence $\bm f \in F_k(t)$, which is associated with a specific pair of $i_p$ and $i_m$ assignments for every vehicle $i$ in this sequence. 


In view of the complexity of analytically solving \textbf{Problem 1}, even with a fixed  $i_p$ and $i_m$ assignment, we adopt a Model Predictive Control with Control Lyapunov-Barrier Functions (MPC-CLBF) framework similar toearlier work \cite{chen2024optimal}. This framework  integrates (a) A receding horizon optimization approach through Model Predictive Control (MPC)  with (b) Control Lyapunov-Barrier Functions (CLBFs) to guarantee convergence to a safe set in finite time, thus providing an extended stability region compared to the use of standard Control Barrier Functions (CBFs). Thi sis detailed next.

\textbf{CBF and CLBF Transformation of Constraints (\ref{equ: rear-end safety const})-(\ref{equ:VehicleConstraints})}
CBFs are used for three main reasons: (a) The original constraints (\ref{equ: rear-end safety const})-(\ref{equ:VehicleConstraints}) can be replaced by CBF-based constraints that imply (\ref{equ: rear-end safety const})-(\ref{equ:VehicleConstraints}), hence they guarantee their satisfaction, 
(b) the forward invariance property of CBFs guarantees constraint satisfaction over all future times,
(c) the new CBF-based constraints are \emph{linear} in the control, which drastically reduces the computational cost of determining them and enables real-time implementation.
In particular, based on the vehicle dynamics \eqref{VehicleDynamics}, we define $f(\bm x_i(t))=[v_i(t),0]^T, g(\bm x_i(t))=[0,1]^T$. 
Each of the constraints \eqref{equ: rear-end safety const}-\eqref{equ:VehicleConstraints} can be expressed in the form $b_j(\bm x_i(t))\geq 0$ for each of the constraints indexed by $j\in 1,2, \ldots$. The 
CBF method maps $b_j(\bm x_i(t))\geq 0$ into a new constraint 
(details may be found in 
\cite{xiao2023safe}, 
\cite{xu2022decentralized})
which directly involves the control $u_i(t)$ and takes the general form
\begin{equation}
L_{f}b_{j}(\bm x_{i}(t))+L_{g}b_{j}(\bm x_{i}(t))u_{i}(t)+\gamma_j (b_{j}(\bm %
x_{i}(t)))\geq 0 \label{equ:cbf}
\end{equation}%
where $L_{f},L_{g}$ denote the Lie derivatives of $b_{j}(\bm x_{i}(t))$ along $f$ and $g$ respectively and $\gamma_j (\cdot )$ denotes a class-$\mathcal{K}$ function (in practice, a linear class-$\mathcal{K}$ function is often used). 

The price to pay in this transformation is due to the fact that the CBF-based constraint above is a \emph{sufficient} condition for $b_j(\bm x_i(t))\geq 0$ (see \cite{xiao2023safe}) and may, therefore, be conservative, i.e., 
it introduces a trade-off between performance and safety. This trade-off can be controlled through the choice of $\gamma_j (\cdot )$.

However, in our problem that involves multiple interconnected CZs, we face an additional complication in the use of CBFs. In particular,
the forward invariance property relies on the initial satisfaction of the original constraints \eqref{equ: rear-end safety const}-\eqref{equ:VehicleConstraints}. While this may be externally enforced on CAVs prior to their entering a distinct CZ as in prior work, here we are dealing with tightly coupled CZs and this assumption is frequently violated as a vehicle leaves one CZ and immediately enters the next.
To handle the issue of a CAV entering a new CZ without fully satisfying the safety constraints required in this CZ, we employ \emph{Control Lyapunov-Barrier Functions (CLBFs)} instead of basic CBFs. 
By appropriately selecting an \emph{exponential} class-$\mathcal{K}$ function instead of the usual linear one in \eqref{equ:cbf}, CLBFs enable finite-time convergence to a safe set when the system initially lies outside this set \cite{xiao2021high}.

Next, we derive the CBF and CLBF constraints of the form \eqref{equ:cbf} with different class-$\mathcal{K}$ function replacing the original constraints.

\emph{Constraint 1 (Vehicle limitations)}: Let $b_1(\bm{x}_{i}(t))=v_{max} - v_{i}(t)$, $b_2(\bm{x}_{i}(t)) = v_{i}(t) - v_{min}$. Following \eqref{equ:cbf}, the corresponding CBF constraints are:
\begin{equation}\label{eqn: CBF_velocity_1}
    -u_{i}(t) + \gamma_1(b_1(\bm{x}_{i}(t)))\geq 0
\end{equation} 
\begin{equation}\label{eqn: CBF_velocity_2}
    u_{i}(t) + \gamma_2(b_2(\bm{x}_{i}(t)))\geq 0
\end{equation} 

\emph{Constraint 2 (Rear-end safety constraint)}: Let $b_3(\bm{x}_{i}(t))=z_{i,i_p}(t)-\varphi v_{i}(t) -\delta $. Similarly, the corresponding CBF constraint is:
\begin{equation}\label{eqn: CBF_rearend}
    v_{i_p}(t) - v_{i}(t) - \varphi u_{i}(t) + \gamma_3(b_3(\bm{x}_{i}(t))) \geq 0
\end{equation}

\emph{Constraint 3 (Safe merging constraint)}: 
The safe merging constraint \eqref{equ: Merging safe} only applies to specific time instants $t_{i_m}^k$. This poses a technical complication due to the fact that a CBF/CLBF must always be in a continuously differentiable form.  We can convert (\ref{equ: Merging safe}) to a continuous time form using the technique in \cite{xiao2021decentralized} to obtain:
\begin{equation}\label{eqn:safe_merging_convert}
z_{i,i_m}(t)\geq \frac{\varphi\cdot x_{i_m}(t)}{L_{i_m}} v_{i}(t)+\delta,  ~~\forall t\in [t_i^{k,0}, t_{i_m}^k]   
\end{equation} 
where $t_i^{k,0}$ denotes the time CAV $i$ enters the road segment directly towards $M_k$ and $t_{i_m}^k$ denotes the time CAV $i_m$ reaches $M_k$. Note that the boundary condition in \eqref{eqn:safe_merging_convert} at $t=t_{i_m}^k$ when $x_{i_m}(t) = L_{i_m}$ is exactly \eqref{equ: Merging safe}, while for $t\in [t_i^{k,0}, t_{i_m}^k)$ the constraint is inactive.

An unsafe initial state frequently arises when the merging safety constraint (\ref{equ: Merging safe}) is violated.
This typically happens when a vehicle enters the CZ and immediately becomes a conflicting CAV $i_m$ of some existing CAV $i$, particularly when the relative distance from the current position of CAV $i$ to the next MP is comparable to that of CAV $i_m$ to the same MP. 
Although the continuous form \eqref{eqn:safe_merging_convert} is used for the safe merging constraint, its satisfaction is only needed at the MP. Thus, if \eqref{eqn:safe_merging_convert} is violated before reaching the MP, CAVs $i$ and $i_m$ still have time to adjust their states and approach the safe set. In such cases, we can use a CLBF as explained next.
Let 
\begin{align}\label{eqn: b_4 merging}
    b_4(\bm{x}_{i}(t))=(L_i-x_{i}(t)) - &(L_{i_m} - x_{i_m}(t))\nonumber\\&-\frac{\varphi}{L_{i_m}}x_{i_m}(t)\cdot v_{i}(t) -\delta
\end{align}
from which we derive the CLBF constraint:
\begin{align}\label{eqn:clbf_safe_merging_1}
    L_{f}b_4(\bm x_{i}(t))+L_{g}&b_4(\bm x_{i}(t))u_{i}(t) + p\cdot sign(b_4(\bm x_{i}(t)) \nonumber\\ & \cdot |b_4(\bm x_{i}(t))|^q \geq 0, 
    ~~\forall t\in [t_i^{k,0}, t_{i_m}^k]  
\end{align}
The $sign$ and absolute value functions are used to prevent $p\cdot(b_4(\bm x_{i}(t)))^q$ from being an imaginary number when $b_4(\bm{x}_{i}(t))<0$.

Thus, when $b_4(\bm x_{i}(t))<0$ which indicates that the state is not in the safe set, we have the CLBF-based condition derived from the safe merging constraint:
\begin{align}\label{eqn:clbf_safe_merging_2}
    v_{i_m}(t)&-v_i(t)-\frac{\varphi}{L_{i_m}}x_{i_m}(t)u_i(t) -\frac{\varphi}{L_{i_m}}v_{i_m}(t)v_i(t)\nonumber \\ 
    & - p \cdot|L_i -x_i(t) - L_{i_m} +x_{i_m}(t) \\ \nonumber
    & \quad -\frac{\varphi}{L_{i_m}}x_{i_m}(t)v_i(t)-\delta|^q \geq 0, ~~\forall t\in [t_i^{k,0}, t_{i_m}^k]
\end{align}
where the parameters $p$ and $q$ can be appropriately selected to guarantee the finite-time convergence of the CLBF (see Proposition 1 in \cite{chen2024optimal}).

\emph{Constraint 4 (Lateral Safety Constraint)} Similarly, letting $b_5(\bm{x}_{i}(t))=w_h\cdot g -\kappa_i(d_{i}(t))\cdot v_{i}(t)^2\cdot h $, the corresponding CBF constraint is:
\begin{equation}\label{eqn: CBF_lateral}
    -2\kappa_i(d_{i}(t)) h v_{i}(t) u_{i}(t)  + \gamma_5(b_5(\bm{x}_{i}(t))) \geq 0
\end{equation}

To summarize, by applying the CBF/CLBF transformation to replace the original constraints, we leverage its forward invariance property to efficiently compute an optimal control solution 
of the original problem (\ref{eqn:obj1}) with the constraints 
(\ref{equ: rear-end safety const})-(\ref{equ:VehicleConstraints}) replaced by their associated CBF-based constraints above. This solution is suboptimal relative to (\ref{eqn:obj1}) but it ensures safety and is relatively simple to obtain online as shown in \cite{chen2024optimal}
for the case where all vehicles are CAVs.

\begin{figure}
    \centering
    \includegraphics[width=0.9\columnwidth]{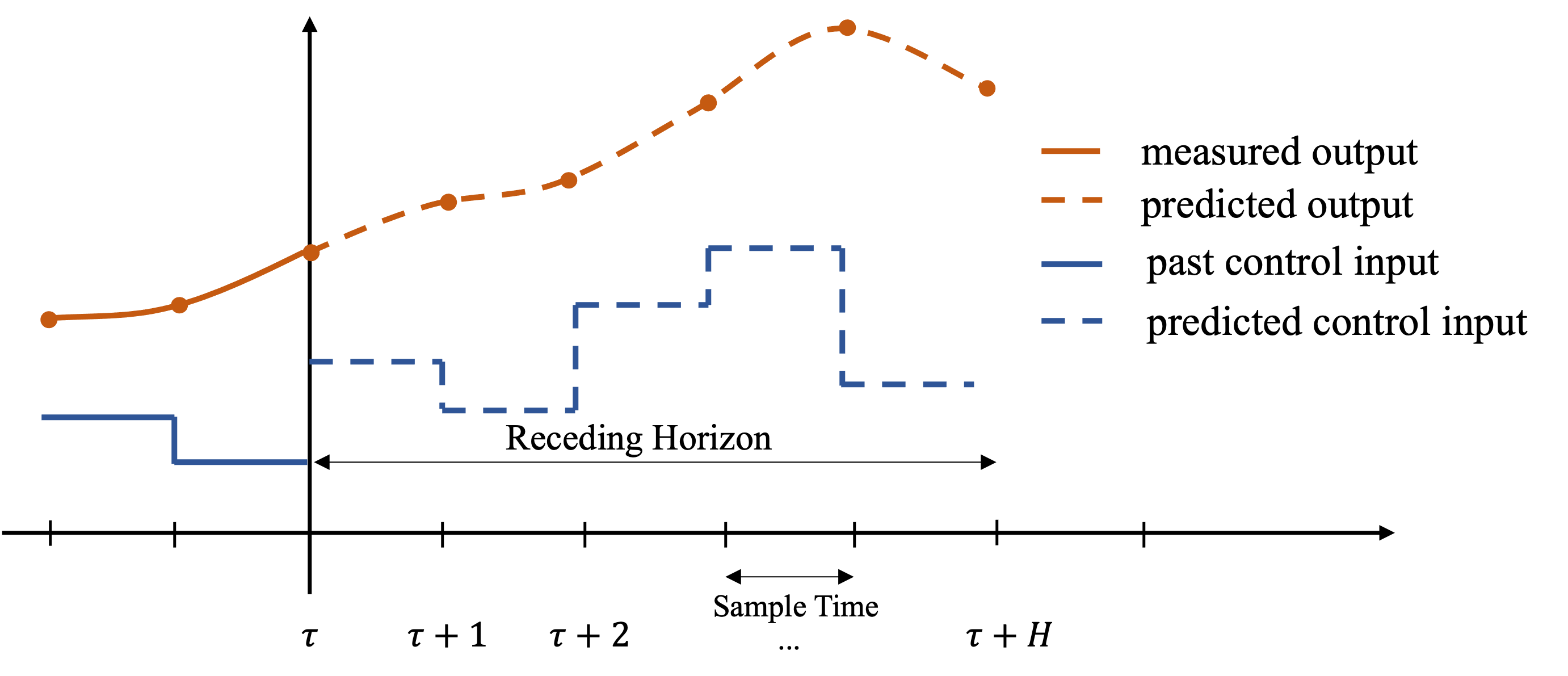}
    \caption{Illustration of MPC framework}
    \label{fig:MPC_demo}
\end{figure}

\textbf{MPC-based Optimal Control Problem Solution}
Given the roundabout's compact geometry which couples closely-spaced MPs, control actions propagate rapidly across different CZs, and the ``myopic'' control determined at a specific time cannot ensure that all constraints remain satisfied in the future especially when a vehicle is changing CZ (e.g., they may conflict with the control constraints in (\ref{equ:VehicleConstraints})). 
In addition, the continuously varying curvature along the entire roundabout route makes it more challenging for myopic control to maintain future safety constraints.
To overcome these challenges, we use a receding-horizon MPC framework to evaluate future performance over a horizon of $H$ lookahead steps, as shown in Fig.\ref{fig:MPC_demo}. 
The MPC method predicts future interaction with a dynamic environment and other vehicles, especially when changing CZs, enabling proactive adjustments to achieve optimality over a tunable receding horizon $H$, thus effectively mitigating the ``myopic'' issue discussed earlier.

Firstly, we discretize time into equal intervals of length $T_d$. At current time $t$, corresponding to time step $\tau$ ($t = \tau\cdot T_d$), the dynamics are discretized such that $\bm{x}_{i,\tau,h} = \bm{x}_i(t+h\cdot T_d)$, $d_{i,\tau,h} = d_i(t+h\cdot T_d)$,  where $h = 0,1, 2,\ldots$ is the $h$-th time step from $t$. 
During each interval $[hT_d, (h+1)T_d]$, we assume the  control input $u_{i,\tau,h}$ is held constant at its value at the start of the interval.
We then set the control, velocity, position trajectory over the next $H$ time steps as a vector respectively: $\bm{u}_{i,\tau}=[u_{i,\tau,0},\ldots u_{i,\tau,H-1}]$, $\bm{v}_{i,\tau}=[v_{i,\tau,1},\ldots v_{i,\tau,H}]$ and $\bm{x}_{i,\tau}=[x_{i,\tau,1},\ldots x_{i,\tau,H}]$. 
When adopting sequence $\bm f$, the corresponding trajectories are denoted as $\bm{u}_{i,\tau}(\bm f)$, $\bm v_{i,\tau}(\bm f)$ and $\bm x_{i,\tau}(\bm f)$.
We also denote the performance of CAV $i$ over the next $H$ time steps from time point $\tau$ under a given sequence $\bm f$ as $J_{i,\tau}^H(\bm f)$. For simplicity, we omit the subscript $\tau$ hereafter.
Note that in this discretization process, the travel time ($t_i^f-t_i^0$) of CAV $i$ in \eqref{eqn:obj1} can no longer be directly minimized  within a fixed time interval. Thus, we instead minimize the speed deviation of CAV $i$ from its desired speed. Additionally, we normalize each term to balance the contributions of energy, speed, and comfort.

The MPC formulation of the optimal control problem comprising the CBF and CLBF transformation of constraints can be represented by \textbf{Problem 2}:


{\small \begin{align}
\label{eqn: MPC-CLBF with given sequence}
\min_{\bm{u_i}}~J_i^H(\bm f)&=\sum_{h=0}^{H-1}(\frac{u_{i,h}^2}{\max\{ u_{i,max}^2,u_{i,min}^2\}}  \nonumber \\
& ~~~ +\lambda_1 \frac{(v_{i,h} - v_d)^2}{(v_{max}-v_{min})^2}
 +\lambda_2\frac{\kappa_i(d_{i,h})v_{i,h}^2}{\kappa_{max}v_{max}^2}) \nonumber \\
s.t. ~~~  &\eqref{equ:VehicleConstraints}, \eqref{eqn: CBF_velocity_1}, \eqref{eqn: CBF_velocity_2}, \eqref{eqn: CBF_rearend}, \eqref{eqn:clbf_safe_merging_2}, \eqref{eqn: CBF_lateral}\nonumber \\
&~~ t = (\tau + h)\cdot T_d, \nonumber \\
&~~x_{i,0} = x_i(\tau\cdot T_d), ~~x_{i,h+1} = x_{i,h} + T_d\cdot v_{i, h} \nonumber \\
&~~v_{i,0} = v_i(\tau\cdot T_d), ~~v_{i,h+1} = v_{i,h}+ T_d \cdot u_{i, h} \nonumber \\
&~~ \forall h\in \{0,\ldots H-1\}
\end{align}}
where $\lambda_1$ and $\lambda_2$ are the positive weights for speed and discomfort respectively, $v_d$ is a desired speed, and $\kappa_{max}$ is the maximum curvature within the roundabout.
This optimal control problem can then be efficiently tackled by solving the MPC problem at each sequence decision round. 


By summing the calculated performance metrics $J^H_i(\bm f)$ for all CAVs $i \in \bm f\cap S^c_k(t)$, the overall performance of $\bm f$ is obtained, denoted as ${J}_k^H(\bm f)$.

\subsection{Determine Optimal Safe Sequence and Motion Control}
Once we have evaluated the performance of each feasible safe sequence $\bm f \in F_k(t)$ within CZ $k$, the optimal sequence is chosen to be the one with the best performance, i.e., 
\begin{equation}\label{eqn:OptimalSequence}
    \bm{f}_k^*(t) = \text{arg}\min_{\bm f\in F_k(t)} \sum_{i \in \bm f \cap S^c_k(t)}J_{i}^H(\bm f)
\end{equation}
This process is conducted for each CZ $k$, and we denote the performance under an optimal sequence at CZ $k$ as $J_k^{H*}$, where $J_k^{H*} = \sum_{i\in \bm f_k^*(t)}J_i^H(\bm f_k^*(t))$. We can then also obtain the optimal trajectories $\bm{u}_{i}^{*},\bm{v}_{i}^{*}, \bm{x}_{i}^{*}$ for each CAV $i\in \bm f_k^*(t)$.
This is detailed in \textbf{Algorithm \ref{alg: resequence}}.

Optimal safe sequences are updated in an event-driven manner,
ensuring timely re-evaluation of feasible safe sequences and their corresponding motion control trajectories. 
In particular, this re-evaluation is triggered by vehicle entering/exiting events, or CZ changing events, as well as by a predefined timeout event. The timeout mechanism is particularly critical to account for the unpredictable behavior of HDVs, ensuring that the system remains adaptive and responsive to dynamic traffic conditions.

\begin{algorithm}[!ht]
\DontPrintSemicolon
  
  \KwInput{$H$, $S(t), S_k(t), k\in \{1,2,3\}$}
  \KwOutput{$\bm{f}_k^*(t),k\in\{1,2,3\}$; $\bm{u}_{i}^{*},\bm{v}_{i}^{*}, \bm{x}_{i}^{*}, i\in S(t)$ }
  \For{$k \in \{1,2,3\}$}    
    { 
        $J_k^{H*} = \infty$;
        $\bm{f}_k^*(t) = \emptyset$;\\
        Get feasible sequence set $F_k(t)$;\\
        \For{$\bm f \in F_k(t)$}
        {
           $J_k^H(\bm f)=0$;\\
           assign $i_p, i_m$ to $\forall$ CAV $i \in S_k(t)$ based on $\bm f$;\\
           \For{$i\in \bm f$}
           {
               get $J_{i}^H(\bm f), \bm{u}_{i}(\bm f), \bm{v}_{i}(\bm f),\bm{x}_{i}(\bm f)$ by solving \textbf{Problem 2}
               
               $ J_k^H(\bm f) = J_k^H(\bm f) + J_i^H(\bm f)$
          }
          \If{$ J_k^H(\bm f) < J_k^{H*}$}
          {
            $J_k^{H*} = J_k^H(\bm f)$;
            $\bm{f}_k^*(t) = \bm f$;\\
            \For{$i\in \bm{f}$}
            {
                $\bm{u}_i^{*},\bm{v}_i^{*}, \bm{u}_i^{*} = \bm{u}_i(\bm f), \bm{v}_i(\bm f),\bm{x}_i(\bm f) $
            }
          }   
        }        
        }

\caption{Optimal Sequencing over a Roundabout}
\label{alg: resequence}
\end{algorithm}

\section{Simulation Results}\label{Sec: Simulation}
This section presents simulation results that evaluate the performance of CAVs and HDVs in mixed traffic scenarios.
The analysis uses the OSS framework, which combines the SS sequencing policy with motion control derived through the MPC-CLBF setting. A scenario with 100\% HDVs serves as a baseline to highlight the impact of CAV motion control. Additionally, a second baseline is established by implementing a Baseline Sequencing (BS) policy where CAVs on the entry road ($c_i=1$) always yield to HDVs 
on the curve road within the same CZ in the roundabout, similar to typical real-world driving rules.

We evaluate system performance using two categories of metrics: \emph{efficiency} and \emph{safety} metrics. The efficiency metrics quantify system performance in terms of average energy consumption, travel time, and passenger discomfort. Safety metrics, on the other hand, measure potential risks both on the same road and across different roads.
The ``infeasible count'' metric measures the number of times when no sequence is feasible for a merging group by solving \textbf{Problem 2}. This is always due to conflicts between CLBF constraints and control limits when a new vehicle enters the roundabout randomly, making it impossible to construct a feasible CLBF function. 
 The ``unsafe count'' is the number of times when the rear-end safety constraint \eqref{equ: rear-end safety const} is violated by a vehicle, which could happen due to insufficient space between a vehicle and its conflicting vehicle ($i_m$) at the next MP when vehicle $i_m$ just crossed the MP. Note that ``unsafe'' behavior is assessed based on the conservative, speed-dependent rear-end safety constraint, rather than actual physical collisions. In fact, no collision is ever observed during our simulation experiments. 
 Similarly, the ``hard deceleration count'' is the number of times when the minimum allowable deceleration is applied. 
 Due to the compact geometry of the roundabout with closely spaced MPs, the risk of merging conflicts is higher compared to other conflict types.
In order to measure such risk, we adopt the Post-Encroachment Time (PET) metric, a standard quantity for identifying crossing conflicts by measuring the time difference between two cross-traffic vehicles in the conflict area  \cite{alhajyaseen2015integration, paul2020post,songchitruksa2006extreme}. 
A PET threshold of 1 second is recommended in \cite{paul2020post} as the optimal value for identifying critical crossing conflict situations under highly heterogeneous traffic conditions by empirical evidence. 
Based on this, we define a safety metric, ``PET critical count'', which records the number of time steps where the PET exceeds the safe threshold of 1 second.

Using the two baselines and two groups of performance metrics, we conduct several simulation experiments across various traffic demand scenarios.
The basic parameter settings are as follows: $L_k^0=L_k^1=60m, \forall k\in \{1,2,3\}$, $\delta=0m$, $\varphi=1.8s$, $v_{max}=20m/s, ~v_{min}=0m/s, ~u_{i,max}=4m/s^2, ~u_{i,min}=-4m/s^2 $for all vehicle $i$. $\lambda_1=0.3, \lambda_2 = 0.02$.
We model HDV behavior using the Intelligent Driver Model (IDM) \cite{treiber2000congested}, and assume that all HDVs exhibit average driving behavior, i.e., $a_j=0 ~~\forall j\in S^h(t)$ except for the specific comparison section. We consider a \emph{symmetric} layout configuration as shown in Fig. \ref{fig:roundabout}. The simulation time step is 0.1 s, and the total simulation time for traffic generation is 1000 s.

\subsection{Balanced Traffic Demand}\label{subsec: sim_balanced_traffic}




 In this case, the simulated incoming traffic is generated through Poisson processes with all rates set to 396 vehicles/h, with randomly assigned exit points.  

\begin{figure*}
    \centering
    \includegraphics[width=1.8\columnwidth]{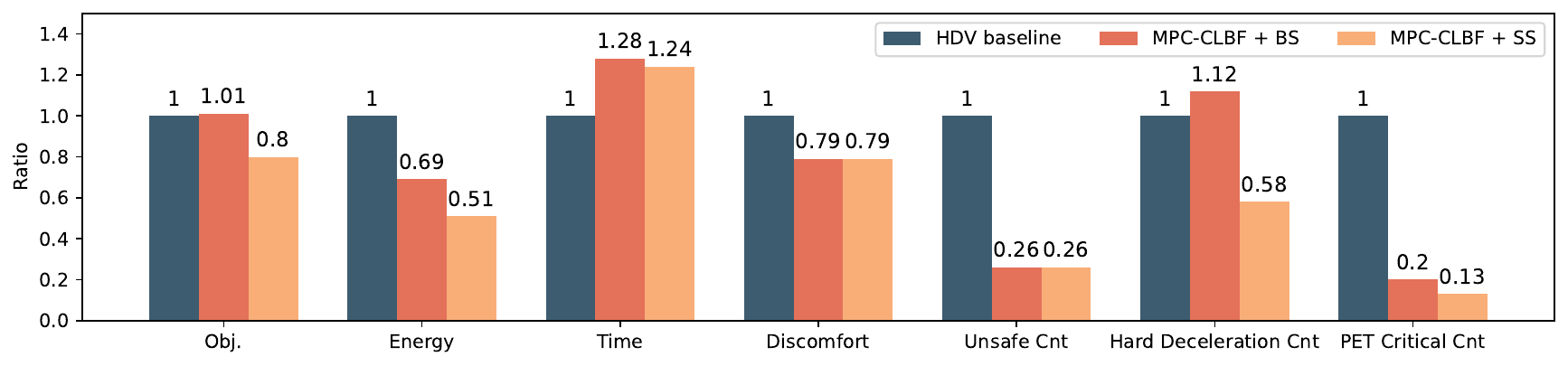}
    \caption{Performance comparison for different policies under CAV penetration rate 0.8}
    \label{fig:cav_08_bar}
\end{figure*}

\textbf{Improvements by Safe Sequencing Policy}
We evaluate the performance of three scenarios: (a) a baseline where all vehicles are HDVs; (b) a mixed traffic scenario with 80\% of vehicles as CAVs controlled by the MPC-CLBF framework, utilizing the Baseline Sequencing (BS) policy introduced earlier; and (c) a similar mixed traffic scenario where 60\% of vehicles are CAVs controlled by the MPC-CLBF framework but with the Safe Sequencing (SS) policy.
The total number of vehicles and their initial states remain consistent across all scenarios, with HDV behavior governed by the same logic in each case. This setup enables us to assess the impact of the MPC-CLBF framework under mixed traffic conditions and to isolate the performance improvements attributable to the sequencing policy. 
The normalized performance metrics 
 are presented in Fig.\ref{fig:cav_08_bar}, where the average metric values for the HDV-only case serve as the baseline and are normalized to 1. The results for the other two cases are shown as ratios relative to this baseline.

The comparative analysis of our OSS framework (MPC-CLBF + SS) in mixed traffic (yellow bars) to the HDV baseline (blue bars) shows notable improvements in vehicle performance and safety in the OSS framework.
Both energy consumption and discomfort are reduced, reflecting smoother vehicle control strategies compared to the abrupt acceleration and deceleration patterns typical of HDVs. However, this improvement comes with a slight increase in travel time. HDVs tend to accelerate aggressively to traverse road segments quickly when no conflicting vehicles are visible, resulting in higher average speeds. Yet, this behavior often leads to sudden deceleration or even collision risks when conflicts arise. As shown in Fig.\ref{fig:cav_08_bar}, all related metrics, including unsafe counts, hard deceleration counts, and critical PET events, show significant improvement in the MPC-CLBF + SS case. This improvement is attributed to our ability to reduce merging conflict risks by limiting HDV behavior through intervention from the front CAV, effectively excluding HDVs from potential merging conflict vehicle pairs.
Thus, the trade-off of slightly longer travel times is justified by the significant safety benefits.

On the other hand, within the same MPC-CLBF framework, the SS policy (yellow bars) demonstrates superior performance compared to the BS policy (orange bars) in both efficiency and safety metrics. 
The BS policy mandates that CAVs on the entry road always yield to HDVs
on the curve road, resulting in more frequent vehicle decelerations and potential complete stops. While this yielding behavior reduces merging collision risk from unpredictable HDV behavior and enhances smoothness of traffic flow on curve roads, it also leads to increased travel time and energy consumption due to repeated deceleration maneuvers. In fact, the BS policy experiences 93.1\% more hard deceleration counts compared to the SS policy. 
Notably, while the BS policy achieves a low unsafe count by reducing conflicts through a clear passing sequence priority, the SS policy attains a similar safety level while consuming less energy and achieving better travel time efficiency.

\begin{figure}[t!]
    \subfloat[BS trajectory \label{subfig:demo_bs}]{
    
        \includegraphics[width=\linewidth]{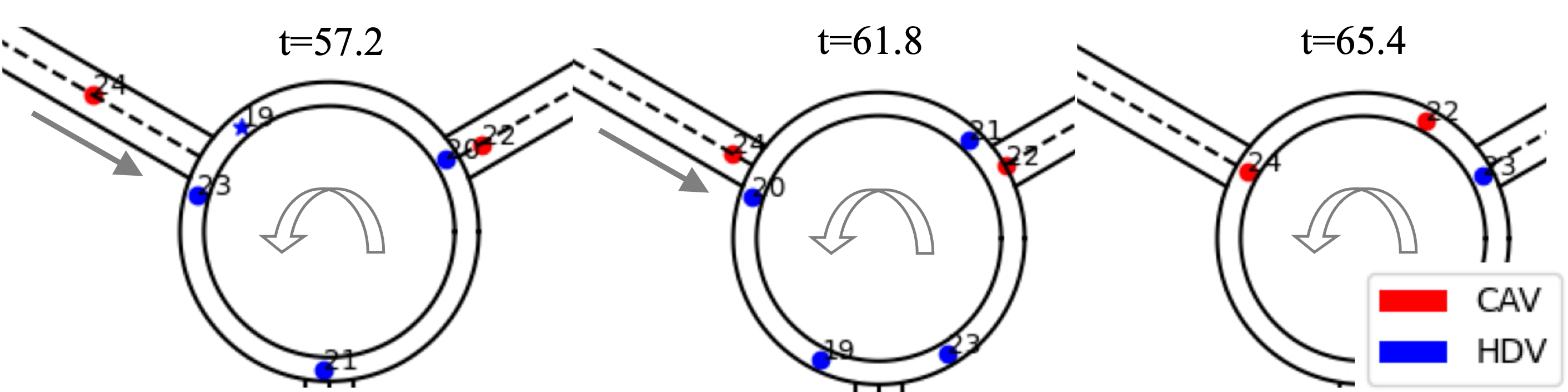}%
    }\\
    \subfloat[SS trajectory \label{subfig:demo_ss}]{%
        \includegraphics[width=\linewidth]{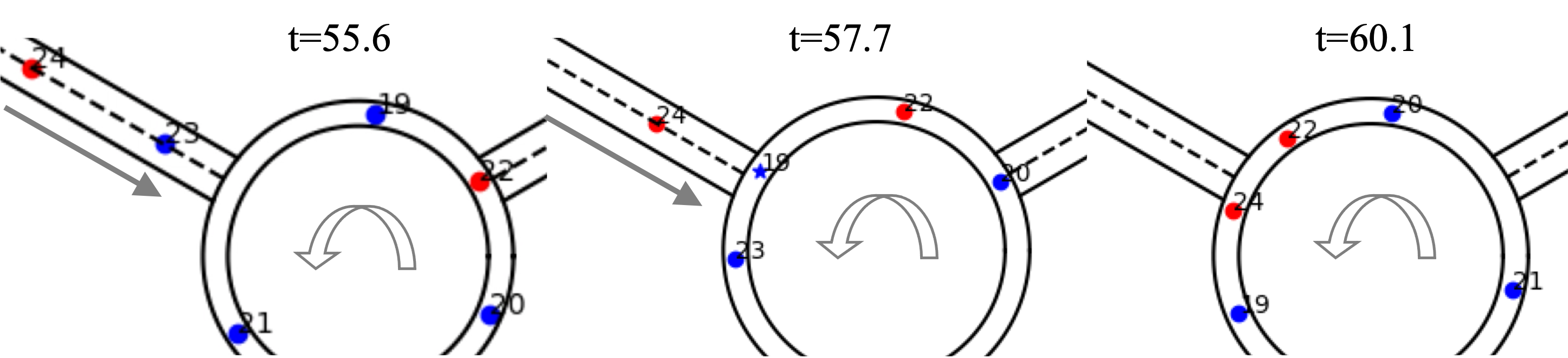}%
        \label{subfig:demo_ss}
    }\\
    \subfloat[$v$ trajectories]{%
        \includegraphics[width=\linewidth]{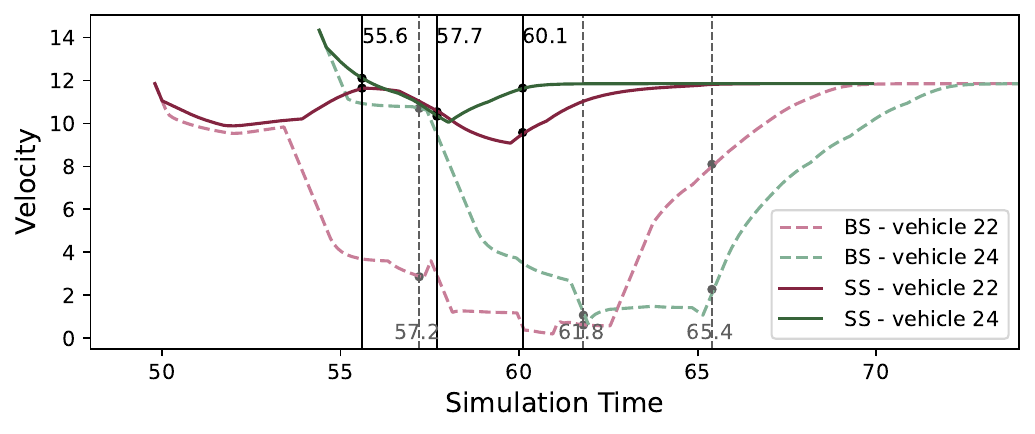}
        \label{subfig:demo_traj_v}%
    }
    \caption{{\small Comparison of trajectories for BS and SS policy: (a) is the partial roundabout state under \emph{BS policy} at time 57.2s, 61.8s, 65.4s; (b) is the partial roundabout state under \emph{SS policy} at time 55.6s, 57.7s, 60.1s. In both plots, red agents indicates CAV while blue agents indicates HDV. Agents with star shape indicates violation of safety constraint. (c) is the velocity trajectories under both policies for vehicle 22 and 24 respectively. }}
    \label{fig:traj demo}
\end{figure}

The distinction between the two policies is further illustrated through the trajectory comparison in Fig. \ref{fig:traj demo}. Fig. \ref{subfig:demo_bs} and \ref{subfig:demo_ss} highlight the differences in vehicle behavior under the BS and the SS policies. Under the BS policy, CAV 22 and 24 yield to multiple HDVs at the curved roads and merge only after most HDVs have passed, resulting in delayed merging. In contrast, under the SS policy, CAV 22 and 24 cooperate to merge earlier while maintaining safety. Fig. \ref{subfig:demo_traj_v} presents the velocity trajectories of CAV 22 and 24, where red and blue lines represent the trajectories of CAV 22 and 24, respectively, and solid and dashed lines represent the results of the MPC-CLBF method using the SS and BS policies, respectively. Vertical solid and dashed lines indicate the corresponding times highlighted in the above example.
The trajectory comparison reveals that the BS policy can lead to hard deceleration, even causing vehicles to stop when reacting to potential conflicts at merging points, as shown by the dashed lines between approximately 55 s and 65 s. This behavior not only wastes energy but also disrupts traffic flow at the entry roads. In contrast, the SS policy mitigates this issue by enabling CAV 22 to predict potential conflicts and decelerate proactively, creating space for CAV 24 to merge safely ahead. 
Additionally, the SS policy gently prompts HDV 20 to decelerate in advance, reducing safety risks. Furthermore, by minimizing unnecessary braking, the SS policy significantly reduces the roundabout traversal time for CAV 22 and 24 compared to the BS policy, demonstrating its efficiency in improving traffic flow and energy conservation.

\textbf{Comparison across Different CAV Penetration Rates.}
Keeping the total number of vehicles and initial states constant, we vary the CAV penetration rate from 0 to 1 in increments of 0.2, while maintaining all other parameter settings unchanged. The performance for CAVs, HDVs, and all vehicles are summarized in Table \ref{table: multi_penetration_rate}.

In terms of efficiency, as the CAV ratio among all vehicles increases, the trends of energy consumption and discomfort diverge from that of travel time. 
Specifically, overall energy consumption decreases with a higher CAV ratio. This trend can be broken down by examining the effects on CAVs and HDVs separately: while the average energy consumption of HDVs fluctuates around a similar level, that of CAVs decreases monotonically, ranging from 49.5\% to 60.8\% of the HDVs' average energy consumption.
This difference arises from the precise control enabled by the MPC-CLBF framework for CAVs, which allows them to predict future interactions with other vehicles and the environment, resulting in smoother driving. 
Therefore, as the CAV ratio increase,  cooperation among CAVs becomes more frequent. Additionally, with fewer HDVs, the need for CAVs to force safe sequencing at MPs by deceleration to intervene HDVs behavior is also reduced. Consequently, less energy is wasted on maintaining safety.
These combined effects lead to a 74.5\% improvement in overall energy efficiency when comparing the full CAV scenario to the full HDV scenario.

On the other hand, the overall travel time follows an inverse trend. As the CAV ratio increases, the travel time for HDVs rises slightly, while CAV travel time initially increases before decreasing, ultimately averaging 19.9\% to 29.9\% higher than that of HDVs. 
This occurs because, at low CAV penetration levels, cooperative interactions between CAVs are infrequent, and CAVs must frequently yield to HDVs to ensure safety, slowing CAV flow. However, as the CAV penetration rate grows, increased cooperation among CAVs helps reduce their travel time.
Meanwhile, HDVs tend to travel faster than CAVs due to their frequent acceleration behavior, as previously explained, resulting in higher average speeds but greater safety risks. As the number of CAVs increases, the average speed within the roundabout decreases, forcing HDVs to slow down slightly as well. This combined effect leads to an overall increase in travel time and a corresponding decrease in discomfort, as discomfort is directly tied to vehicle speed.
In terms of safety, nearly all safety metrics improve for both vehicle types as the CAV ratio increases, though the performance differs between HDVs and CAVs. CAVs consistently demonstrate significantly better average safety metric values compared to HDVs, highlighting their superior control and predictability. 

\begin{table*}[]
\caption{Performance Comparison across Different CAV Penetration Rates}
\huge
\centering
\begin{threeparttable}
\resizebox{\textwidth}{!}{%
\begin{tabular}{|c|ccc|ccc|ccc|ccc|ccc|ccc|}
\hline
\multirow{2}{*}{} &
  \multicolumn{3}{c|}{0} &
  \multicolumn{3}{c|}{0.2} &
  \multicolumn{3}{c|}{0.4} &
  \multicolumn{3}{c|}{0.6} &
  \multicolumn{3}{c|}{0.8} &
  \multicolumn{3}{c|}{1} \\ \cline{2-19} 
 &
  CAV &
  HDV &
  \textbf{all} &
  CAV &
  HDV &
  \textbf{all} &
  CAV &
  HDV &
  \textbf{all} &
  CAV &
  HDV &
  \textbf{all} &
  CAV &
  HDV &
  \textbf{all} &
  CAV &
  HDV &
  \textbf{all} \\ \hline
Avg. Obj.\tnote{1} &
   &
  1.24 &
  \textbf{1.24} &
  1.07 &
  1.29 &
  \textbf{1.24} &
  1.04 &
  1.33 &
  \textbf{1.21} &
  0.97 &
  1.26 &
  \textbf{1.09} &
  0.9 &
  1.29 &
  \textbf{0.99} &
  0.73 &
   &
  \textbf{0.73} \\
Avg. Energy &
   &
  10.94 &
  \textbf{10.94} &
  5.54 &
  10.98 &
  \textbf{9.86} &
  5.14 &
  11.19 &
  \textbf{8.79} &
  4.57 &
  10.45 &
  \textbf{7.13} &
  4.12 &
  10.52 &
  \textbf{5.55} &
  2.79 &
   &
  \textbf{2.79} \\
Avg. Time &
   &
  12.69 &
  \textbf{12.69} &
  16.93 &
  13.16 &
  \textbf{13.93} &
  17.14 &
  13.19 &
  \textbf{14.76} &
  16.75 &
  13.33 &
  \textbf{15.26} &
  16.4 &
  13.68 &
  \textbf{15.79} &
  16.06 &
   &
  \textbf{16.06} \\
Avg. Speed &
   &
  14.54 &
  \textbf{14.54} &
  11.23 &
  14.02 &
  \textbf{13.45} &
  11.2 &
  13.76 &
  \textbf{12.74} &
  11.24 &
  13.65 &
  \textbf{12.29} &
  11.32 &
  13.45 &
  \textbf{11.79} &
  11.49 &
   &
  \textbf{11.49} \\
Avg. Discomfort &
   &
  66.36 &
  \textbf{66.36} &
  51.66 &
  63.44 &
  \textbf{61.02} &
  52.49 &
  60.4 &
  \textbf{57.26} &
  51.05 &
  59.95 &
  \textbf{54.93} &
  49.94 &
  59.73 &
  \textbf{52.13} &
  50.14 &
   &
  \textbf{50.14} \\ \hline
Avg. Unsafe Cnt. &
   &
  7.08 &
  \textbf{7.08} &
  0.34 &
  6.55 &
  \textbf{5.28} &
  0.25 &
  5.79 &
  \textbf{3.59} &
  0.28 &
  4.92 &
  \textbf{2.3} &
  0.8 &
  5.41 &
  \textbf{1.83} &
  0.32 &
   &
  \textbf{0.32} \\
Avg. Hard Deceleration Cnt. &
   &
  3.13 &
  \textbf{3.13} &
  3.42 &
  2.98 &
  \textbf{3.07} &
  2.7 &
  3.08 &
  \textbf{2.93} &
  2.16 &
  2.39 &
  \textbf{2.26} &
  1.63 &
  2.5 &
  \textbf{1.82} &
  0.26 &
   &
  \textbf{0.26} \\
Avg. PET Critical Cnt. &
   &
  0.21 &
  \textbf{0.21} &
  0.02 &
  0.2 &
  \textbf{0.16} &
  0.01 &
  0.16 &
  \textbf{0.1} &
  0 &
  0.17 &
  \textbf{0.07} &
  0 &
  0.12 &
  \textbf{0.03} &
  0 &
   &
  \textbf{0} \\
Avg. Infeasible Cnt. &
   &
   &
  \textbf{} &
  3.42 &
   &
  \textbf{} &
  2.7 &
   &
  \textbf{} &
  2.15 &
   &
  \textbf{} &
  1.62 &
   &
  \textbf{} &
  0.26 &
   &
  \textbf{} \\ \hline
\end{tabular}%
}

\begin{tablenotes}\scriptsize
\item [1] Avg. obj =  Avg. normalized energy + $\lambda_1 \times$ Avg. normalized speed deviation +  $\lambda_2 \times$ avg. normalized discomfort, where $\lambda_1 = 0.3, \lambda_2 = 0.02$. See \eqref{eqn: MPC-CLBF with given sequence}.
\end{tablenotes}
\end{threeparttable}

\label{table: multi_penetration_rate}
\end{table*}

\subsection{Variable Traffic Demand} \label{subsec: sim_various_traffic}
We next evaluate the OSS approach's effectiveness in handling unbalanced and heavy traffic conditions.

\begin{figure*}%
\centering
\begin{subfigure}{.65\columnwidth}
\includegraphics[width=\columnwidth]{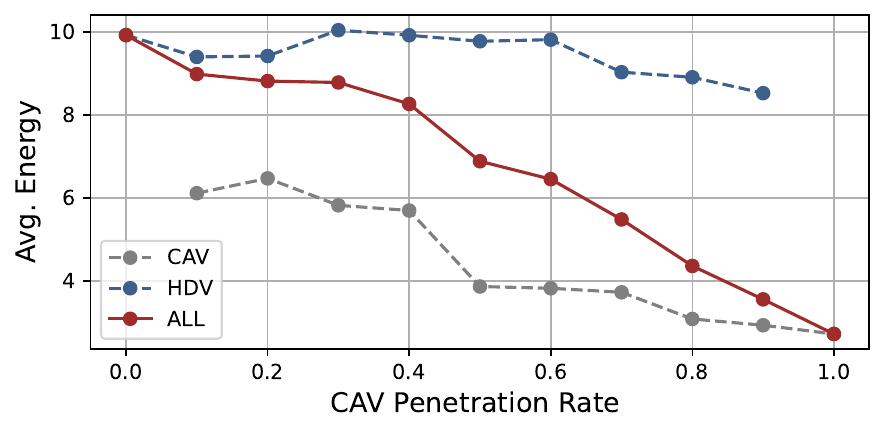}%
\caption{Energy Trajectories}%
\label{subfig: unbalanced_energy}%
\end{subfigure}\hfill%
\begin{subfigure}{.65\columnwidth}
\includegraphics[width=\columnwidth]{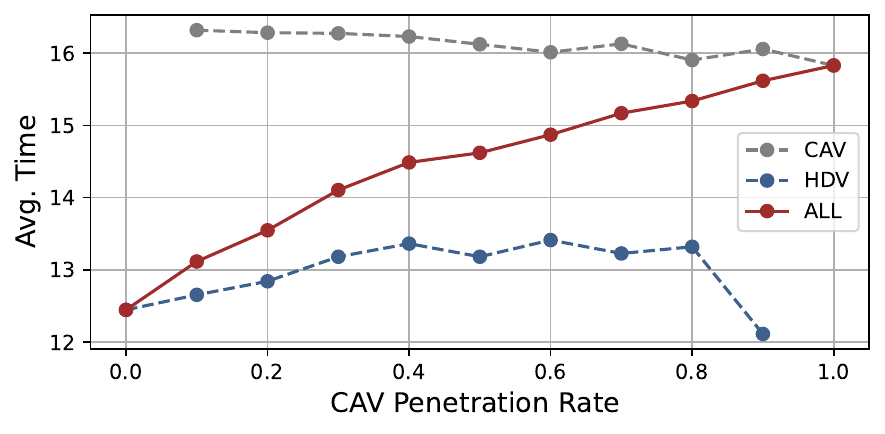}%
\caption{Time Trajectories}%
\label{subfig: unbalanced_time}%
\end{subfigure}\hfill%
\begin{subfigure}{.65\columnwidth}
\includegraphics[width=\columnwidth]{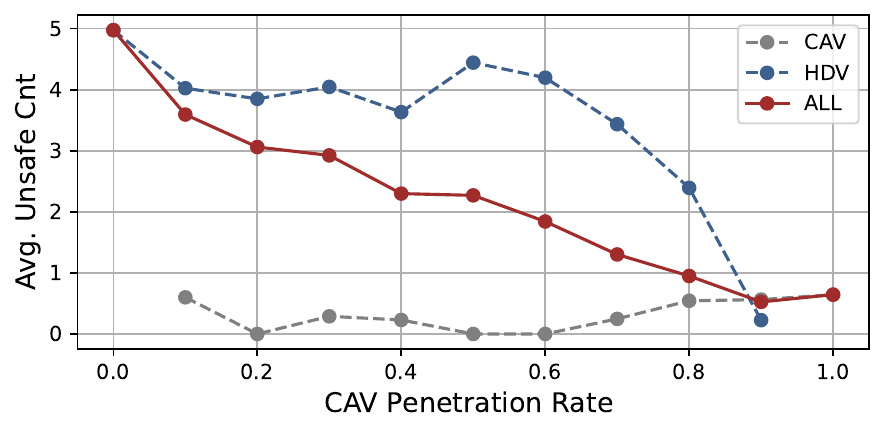}%
\caption{Unsafe Count Trajectories}%
\label{subfig: unbalanced_unsafe}%
\end{subfigure}%
\caption{Comparison of Performance under \emph{Unbalanced} Traffic}
\label{fig: unbalanced}
\end{figure*}

\textbf{Unbalanced traffic demand:} 
In this case, we set the traffic arrival rates to be 108 vehicles/h, 540 vehicles/h, and 540 vehicles/h for $O_1, O_2$ and $O_3$ respectively, so that the total traffic demand level for the whole roundabout is similar to the previous case. All other parameter settings are the same. 
The trajectories of energy consumption, travel time and unsafe count along increasing CAV ratio value for both CAVs only, HDVs only and their combined effects are shown in Fig. \ref{fig: unbalanced}.
The trends align with those observed in the balanced traffic scenario: overall energy consumption and unsafe events improve significantly as CAV ratio increases, reflecting smoother and safer driving behavior, albeit at the cost of a slight increase in travel time.



\begin{figure*}%
\centering
\begin{subfigure}{.65\columnwidth}
\includegraphics[width=\columnwidth]{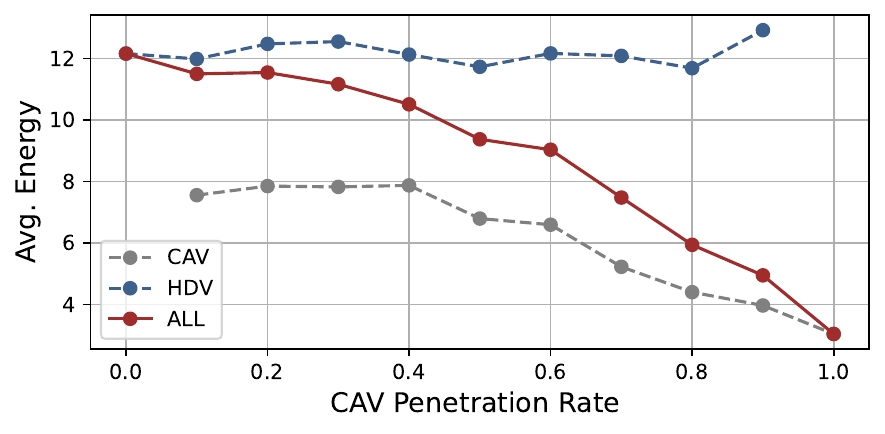}%
\caption{Energy Trajectories}%
\label{subfig: heavy_energy}%
\end{subfigure}\hfill%
\begin{subfigure}{.65\columnwidth}
\includegraphics[width=\columnwidth]{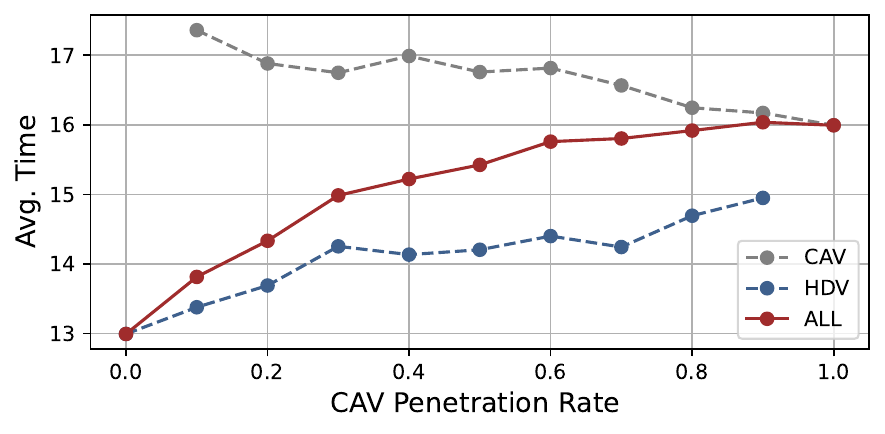}%
\caption{Time Trajectories}%
\label{subfig: heavy_time}%
\end{subfigure}\hfill%
\begin{subfigure}{.65\columnwidth}
\includegraphics[width=\columnwidth]{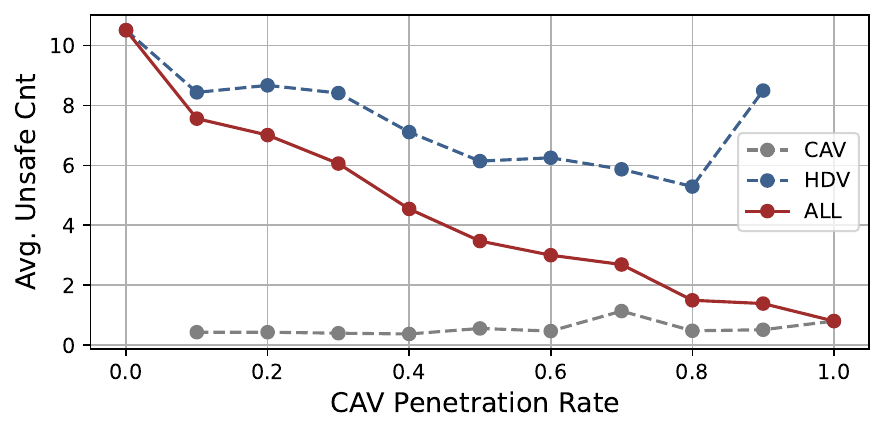}%
\caption{Unsafe Count Trajectories}%
\label{subfig: heavy_unsafe}%
\end{subfigure}%
\caption{Comparison of Performance under \emph{Heavy} Traffic}
\label{fig: heavy}
\end{figure*}

\textbf{Heavy traffic demand:} 
To evaluate the performance of the MPC-CLBF method under mixed and heavy traffic conditions, we increased the arrival rates at all origins to 576 vehicles/hour, raising total traffic volume by 45.5\% compared to earlier cases to simulate congestion. All other parameters remained unchanged, and the results are presented in Fig. \ref{fig: heavy}.
Under heavy traffic, total travel time, energy consumption, and unsafe events increase overall compared to previous scenarios. However, the improvements in energy consumption and unsafe events become even more pronounced as the CAV penetration rate increases, highlighting the added benefits of CAVs in guiding and smoothing traffic under congested conditions.

The performance under different traffic demands indicates that our Optimal Safe Sequencing framework consistently showcases superiority in simultaneously enhancing efficiency, safety, and robustness.

\subsection{Computational Complexity Analysis}
Let the number of CZs in a roundabout be $N$ and the number of vehicles on the roundabout and entry road segments of CZ $k\in\{1,\ldots N\}$ be $N_k^0$ and $N_k^1$ respectively. Given the assumption of no vehicle overtaking, the number of feasible sequences in CZ $k$ is $\binom{N_k^0+N_k^1}{N_k^0}$. However, the SS policy, which restricts CAVs from merging ahead of an HDV, reduces the feasible sequence set.
In our simulations (Sec. \ref{Sec: Simulation}), with a CAV penetration rate of 0.5, the average number of feasible safe sequences per merging group is 1.17 for balanced normal traffic (396 vehicles/h) and 1.589 for heavy traffic (576 vehicles/h).
The average number of times that \textbf{Problem 2} is solved within a single resequencing calculation is 1.71 for normal traffic and 3.26 for heavy traffic, increasing with higher CAV penetration rates due to more CAVs being involved.  This solution of \textbf{Problem 2} provides motion control for a single CAV in future steps under a given sequence. The average computation time for a single solution is around $100 ms$ on an Intel Core m5 with two 1.2GHz cores using Gurobi as a numerical solver.
Consequently, the total computation time for sequence and motion control updates across all CAVs in the roundabout is approximately $140 ms$ for balanced normal traffic and $285 ms$ for heavy traffic.
Moreover, parallel computing can be employed in practice to evaluate different sequences simultaneously, offering significant computational savings. These results suggest that the Optimal Safe Sequencing (OSS) framework is computationally feasible for real-time implementation.

\section{Conclusion and Future Work} \label{sec: conclusion}
We have presented a control framework for CAVs navigating a single-lane roundabout in mixed traffic, where both CAVs and HDVs coexist without relying on assumptions about HDV behavior. Our proposed Optimal Safe Sequencing (OSS) framework integrates a Safe Sequencing policy with an MPC-CLBF approach, jointly optimizing vehicle sequencing and motion control to balance speed, energy efficiency, and comfort while ensuring safety.
Simulation results demonstrate that safety and energy metrics considerably improve at the expense of modest time increases compared to a fully HDV-controlled scenario and a Baseline Sequencing policy using the same MPC-CLBF structure. Additionally, comparisons across varying CAV penetration rates highlight the benefits of increased CAV presence in the traffic system.
Future work will focus on enhancing CAV decision-making involving HDVs within the roundabout by leveraging learned HDV behavior from past trajectories.

\bibliographystyle{IEEEtran}
\bibliography{ref}

\begin{thebibliography}{10}
\providecommand{\url}[1]{#1}
\csname url@samestyle\endcsname
\providecommand{\newblock}{\relax}
\providecommand{\bibinfo}[2]{#2}
\providecommand{\BIBentrySTDinterwordspacing}{\spaceskip=0pt\relax}
\providecommand{\BIBentryALTinterwordstretchfactor}{4}
\providecommand{\BIBentryALTinterwordspacing}{\spaceskip=\fontdimen2\font plus
\BIBentryALTinterwordstretchfactor\fontdimen3\font minus \fontdimen4\font\relax}
\providecommand{\BIBforeignlanguage}[2]{{%
\expandafter\ifx\csname l@#1\endcsname\relax
\typeout{** WARNING: IEEEtran.bst: No hyphenation pattern has been}%
\typeout{** loaded for the language `#1'. Using the pattern for}%
\typeout{** the default language instead.}%
\else
\language=\csname l@#1\endcsname
\fi
#2}}
\providecommand{\BIBdecl}{\relax}
\BIBdecl

\bibitem{chang2017there}
Y.~S. Chang, Y.~J. Lee, and S.~S.~B. Choi, ``Is there more traffic congestion in larger cities?-scaling analysis of the 101 largest us urban centers,'' \emph{Transport Policy}, vol.~59, pp. 54--63, 2017.

\bibitem{rios2016automated}
J.~Rios-Torres and A.~A. Malikopoulos, ``Automated and cooperative vehicle merging at highway on-ramps,'' \emph{IEEE Trans. on Intelligent Transp. Systems}, vol.~18, no.~4, pp. 780--789, 2016.

\bibitem{liu2023}
S.~Liu, J.~P. Petitti, Y.~Lei, Y.~Liu, and C.~A. Shue, ``By your command: Extracting the user actions that create network flows in android,'' in \emph{14th International Conf. on Network of the Future}, 2023, pp. 118--122.

\bibitem{campi2023roundabouts}
E.~Campi, G.~Mastinu, G.~Previati, L.~Studer, and L.~Uccello, ``Roundabouts: Traffic simulations of connected and automated vehicles—a state of the art,'' \emph{IEEE Trans. on Intelligent Transp. Systems}, pp. 1--21, 2023.

\bibitem{milanes2010automated}
V.~Milan{\'e}s, J.~Godoy, J.~Villagr{\'a}, and J.~P{\'e}rez, ``Automated on-ramp merging system for congested traffic situations,'' \emph{IEEE Trans. on Intelligent Transp. Systems}, vol.~12, no.~2, pp. 500--508, 2010.

\bibitem{xiao2021decentralized}
W.~Xiao and C.~G. Cassandras, ``Decentralized optimal merging control for connected and automated vehicles with safety constraint guarantees,'' \emph{Automatica}, vol. 123, p. 109333, 2021.

\bibitem{armijos2024cooperative}
A.~S.~C. Armijos, A.~Li, C.~G. Cassandras, Y.~K. Al-Nadawi, H.~Araki, B.~Chalaki, E.~Moradi-Pari, H.~N. Mahjoub, and V.~Tadiparthi, ``Cooperative energy and time-optimal lane change maneuvers with minimal highway traffic disruption,'' \emph{Automatica}, vol. 165, p. 111651, 2024.

\bibitem{li2023cooperative}
A.~Li, A.~S.~C. Armijos, and C.~G. Cassandras, ``Cooperative lane changing in mixed traffic can be robust to human driver behavior,'' in \emph{62nd IEEE Conf. on Decision and Control}, 2023, pp. 5123--5128.

\bibitem{bichiou2018developing}
Y.~Bichiou and H.~A. Rakha, ``Developing an optimal intersection control system for automated connected vehicles,'' \emph{IEEE Trans. on Intelligent Transp. Systems}, vol.~20, no.~5, pp. 1908--1916, 2018.

\bibitem{zhang2019decentralized}
Y.~Zhang and C.~G. Cassandras, ``Decentralized optimal control of connected automated vehicles at signal-free intersections including comfort-constrained turns and safety guarantees,'' \emph{Automatica}, vol. 109, p. 108563, 2019.

\bibitem{xu_scaling_2023}
K.~Xu and C.~G. Cassandras, ``Scaling up the optimal safe control of connected and automated vehicles to a traffic network: A hierarchical framework of modular control zones,'' in \emph{26th IEEE Intelligent Transp. Systems Conf.}, 2023, pp. 1448--1453.

\bibitem{rodegerdts2010roundabouts}
L.~A. Rodegerdts, \emph{Roundabouts: An informational guide}.\hskip 1em plus 0.5em minus 0.4em\relax Transp. Research Board, 2010, vol. 672.

\bibitem{bakibillah_bi-level_2021}
A.~Bakibillah, M.~A.~S. Kamal, C.~P. Tan, S.~Susilawati, T.~Hayakawa, and J.-i. Imura, ``Bi-level coordinated merging of connected and automated vehicles at roundabouts,'' \emph{Sensors}, vol.~21, no.~19, p. 6533, 2021.

\bibitem{bichiou_developing_2019}
Y.~Bichiou and H.~A. Rakha, ``Developing an optimal intersection control system for automated connected vehicles,'' \emph{IEEE Trans. on Intelligent Transp. Systems}, vol.~20, no.~5, pp. 1908--1916, 2018.

\bibitem{zhao_optimal_2018}
L.~Zhao, A.~Malikopoulos, and J.~Rios-Torres, ``Optimal control of connected and automated vehicles at roundabouts: An investigation in a mixed-traffic environment,'' \emph{IFAC-PapersOnLine}, vol.~51, no.~9, pp. 73--78, 2018.

\bibitem{xu2021decentralized}
K.~Xu, C.~G. Cassandras, and W.~Xiao, ``Decentralized time and energy-optimal control of connected and automated vehicles in a roundabout,'' in \emph{Proc. of 24th IEEE Intelligent Transp. Systems Conf.}, 2021, pp. 681--686.

\bibitem{garcia_cuenca_autonomous_2019}
L.~García~Cuenca, E.~Puertas, J.~Fernandez~Andrés, and N.~Aliane, ``\BIBforeignlanguage{en}{Autonomous {Driving} in {Roundabout} {Maneuvers} {Using} {Reinforcement} {Learning} with {Q}-{Learning}},'' \emph{\BIBforeignlanguage{en}{Electronics}}, vol.~8, no.~12, p. 1536, 2019.

\bibitem{capasso_intelligent_2020}
A.~P. Capasso, G.~Bacchiani, and D.~Molinari, ``\BIBforeignlanguage{en}{Intelligent {Roundabout} {Insertion} using {Deep} {Reinforcement} {Learning}},'' in \emph{\BIBforeignlanguage{en}{Proc. of 12th {Intl.} {Conf.} on {Agents} and {Artificial} {Intelligence}}}, 2020, pp. 378--385.

\bibitem{mohebifard2022trajectory}
R.~Mohebifard and A.~Hajbabaie, ``Trajectory control in roundabouts with a mixed fleet of automated and human-driven vehicles,'' \emph{Computer-Aided Civil and Infrastructure Engineering}, vol.~37, no.~15, pp. 1959--1977, 2022.

\bibitem{ferrarotti2024autonomous}
L.~Ferrarotti, M.~Luca, G.~Santin, G.~Previati, G.~Mastinu, M.~Gobbi, E.~Campi, L.~Uccello, A.~Albanese, P.~Zalaya \emph{et~al.}, ``Autonomous and human-driven vehicles interacting in a roundabout: A quantitative and qualitative evaluation,'' \emph{IEEE Access}, 2024.

\bibitem{wang2020game}
M.~Wang, N.~Mehr, A.~Gaidon, and M.~Schwager, ``Game-theoretic planning for risk-aware interactive agents,'' in \emph{2020 IEEE/RSJ International Conference on Intelligent Robots and Systems (IROS)}.\hskip 1em plus 0.5em minus 0.4em\relax IEEE, 2020, pp. 6998--7005.

\bibitem{burger2022interaction}
C.~Burger, J.~Fischer, F.~Bieder, {\"O}.~{\c{S}}. Ta{\c{s}}, and C.~Stiller, ``Interaction-aware game-theoretic motion planning for automated vehicles using bi-level optimization,'' in \emph{2022 IEEE 25th International Conference on Intelligent Transportation Systems (ITSC)}.\hskip 1em plus 0.5em minus 0.4em\relax IEEE, 2022, pp. 3978--3985.

\bibitem{wang2021game}
M.~Wang, Z.~Wang, J.~Talbot, J.~C. Gerdes, and M.~Schwager, ``Game-theoretic planning for self-driving cars in multivehicle competitive scenarios,'' \emph{IEEE Transactions on Robotics}, vol.~37, no.~4, pp. 1313--1325, 2021.

\bibitem{sadigh2018planning}
D.~Sadigh, N.~Landolfi, S.~S. Sastry, S.~A. Seshia, and A.~D. Dragan, ``Planning for cars that coordinate with people: leveraging effects on human actions for planning and active information gathering over human internal state,'' \emph{Autonomous Robots}, vol.~42, pp. 1405--1426, 2018.

\bibitem{zhang2023social}
L.~Zhang, Y.~Dong, H.~Farah, and B.~van Arem, ``Social-aware planning and control for automated vehicles based on driving risk field and model predictive contouring control: Driving through roundabouts as a case study,'' in \emph{2023 IEEE International Conference on Systems, Man, and Cybernetics (SMC)}.\hskip 1em plus 0.5em minus 0.4em\relax IEEE, 2023, pp. 3297--3304.

\bibitem{mahdinia2021integration}
I.~Mahdinia, A.~Mohammadnazar, R.~Arvin, and A.~J. Khattak, ``Integration of automated vehicles in mixed traffic: Evaluating changes in performance of following human-driven vehicles,'' \emph{Accident Analysis \& Prevention}, vol. 152, p. 106006, 2021.

\bibitem{zhuo2023evaluation}
J.~Zhuo and F.~Zhu, ``Evaluation of platooning configurations for connected and automated vehicles at an isolated roundabout in a mixed traffic environment,'' \emph{Journal of Intelligent and Connected Vehicles}, vol.~6, no.~3, pp. 136--148, 2023.

\bibitem{chen2024optimal}
Y.~Chen, C.~G. Cassandras, and K.~Xu, ``Optimal sequencing and motion control in a roundabout with safety guarantees,'' in \emph{63rd IEEE Conf. on Decision and Control}, 2024, pp. 5699--5704.

\bibitem{xiao2020decentralized}
W.~Xiao, C.~G. Cassandras, and C.~Belta, ``Decentralized optimal control in multi-lane merging for connected and automated vehicles,'' in \emph{2020 IEEE 23rd International Conference on Intelligent Transportation Systems (ITSC)}.\hskip 1em plus 0.5em minus 0.4em\relax IEEE, 2020, pp. 1--6.

\bibitem{xu2022general}
H.~Xu, W.~Xiao, C.~G. Cassandras, Y.~Zhang, and L.~Li, ``A general framework for decentralized safe optimal control of connected and automated vehicles in multi-lane signal-free intersections,'' \emph{IEEE Transactions on Intelligent Transportation Systems}, vol.~23, no.~10, pp. 17\,382--17\,396, 2022.

\bibitem{polack2017kinematic}
P.~Polack, F.~Altch{\'e}, B.~d'Andr{\'e}a Novel, and A.~de~La~Fortelle, ``The kinematic bicycle model: A consistent model for planning feasible trajectories for autonomous vehicles?'' in \emph{2017 IEEE intelligent vehicles symposium (IV)}.\hskip 1em plus 0.5em minus 0.4em\relax IEEE, 2017, pp. 812--818.

\bibitem{rucco2015efficient}
A.~Rucco, G.~Notarstefano, and J.~Hauser, ``An efficient minimum-time trajectory generation strategy for two-track car vehicles,'' \emph{IEEE Transactions on Control Systems Technology}, vol.~23, no.~4, pp. 1505--1519, 2015.

\bibitem{xu2022decentralized}
K.~Xu, C.~G. Cassandras, and W.~Xiao, ``Decentralized time and energy-optimal control of connected and automated vehicles in a roundabout with safety and comfort guarantees,'' \emph{IEEE Trans. on Intelligent Transp. Systems}, vol.~24, no.~1, pp. 657--672, 2022.

\bibitem{Vogel2003}
K.~Vogel, ``A comparison of headway and time to collision as safety indicators,'' \emph{Accident analysis \& prevention}, vol.~35, no.~3, pp. 427--433, 2003.

\bibitem{sabouni_merging_2023}
E.~Sabouni, H.~S. Ahmad, C.~G. Cassandras, and W.~Li, ``Merging control in mixed traffic with safety guarantees: a safe sequencing policy with optimal motion control,'' in \emph{2023 IEEE 26th International Conference on Intelligent Transportation Systems (ITSC)}.\hskip 1em plus 0.5em minus 0.4em\relax IEEE, 2023, pp. 4260--4265.

\bibitem{chandra2022game}
R.~Chandra, M.~Wang, M.~Schwager, and D.~Manocha, ``Game-theoretic planning for autonomous driving among risk-aware human drivers,'' in \emph{2022 International Conference on Robotics and Automation (ICRA)}.\hskip 1em plus 0.5em minus 0.4em\relax IEEE, 2022, pp. 2876--2883.

\bibitem{xiao2023safe}
W.~Xiao, C.~G. Cassandras, and C.~Belta, \emph{Safe Autonomy with Control Barrier Functions: Theory and Applications}.\hskip 1em plus 0.5em minus 0.4em\relax Springer Nature, 2023.

\bibitem{xiao2021high}
W.~Xiao, C.~A. Belta, and C.~G. Cassandras, ``High order control lyapunov-barrier functions for temporal logic specifications,'' in \emph{2021 IEEE American Control Conf.}, 2021, pp. 4886--4891.

\bibitem{alhajyaseen2015integration}
W.~K. Alhajyaseen, ``The integration of conflict probability and severity for the safety assessment of intersections,'' \emph{Arabian Journal for Science and Engineering}, vol.~40, pp. 421--430, 2015.

\bibitem{paul2020post}
M.~Paul and I.~Ghosh, ``Post encroachment time threshold identification for right-turn related crashes at unsignalized intersections on intercity highways under mixed traffic,'' \emph{International journal of injury control and safety promotion}, vol.~27, no.~2, pp. 121--135, 2020.

\bibitem{songchitruksa2006extreme}
P.~Songchitruksa and A.~P. Tarko, ``The extreme value theory approach to safety estimation,'' \emph{Accident Analysis \& Prevention}, vol.~38, no.~4, pp. 811--822, 2006.

\bibitem{treiber2000congested}
M.~Treiber, A.~Hennecke, and D.~Helbing, ``Congested traffic states in empirical observations and microscopic simulations,'' \emph{Physical review E}, vol.~62, no.~2, p. 1805, 2000.

\end{thebibliography}

\end{document}